\documentstyle[12pt,aaspp4]{article}
\newcommand\etal{{\it et al}. }
\newcommand\be {\begin{equation}}
\newcommand\en{\end{equation}}

\slugcomment{Submitted to {\it The Astrophysical Journal}}

\begin{document}

\title{NONLINEAR EFFECTS IN MODELS OF THE GALAXY:\\ 1. MIDPLANE STELLAR ORBITS 
IN THE PRESENCE OF 3D SPIRAL ARMS}

\author{B\'arbara Pichardo \altaffilmark{1}, Marco Martos \altaffilmark{1}, Edmundo Moreno \altaffilmark{1}, Julia Espresate \altaffilmark{1}}

\altaffiltext{1}{Instituto de Astronom\'\i a, Universidad Nacional Aut\'onoma de 
M\'exico, A.P. 70-264, 04510 M\'exico D.F., M\'exico; Electronic mail: 
barbara@astroscu.unam.mx; marco@astroscu.unam.mx; edmundo@astroscu.unam.mx; julia@astroscu.unam.mx}

\begin{abstract}
With the aim of studying the nonlinear stellar and gaseous response to
the gravitational potential of a galaxy such as the Milky Way, we have
modeled 3D galactic spiral arms as a superposition of inhomogeneous
oblate spheroids and added their contribution to an axisymmetric model
of the Galactic mass distribution. Three spiral loci are proposed
here, based in different sets of observations. A comparison of our
model with a tight-winding approximation shows that the
self-gravitation of the whole spiral pattern is important in the
middle and outer galactic regions. A preliminary self-consistency
analysis taking ${\Omega}_{p}$ = 15 and 20~km~s$^{-1}$~kpc$^{-1}$ for
the angular speed of the spiral pattern, seems to favor the value
${\Omega}_{p}$ = 20~km~s$^{-1}$~kpc$^{-1}$. As a first step to full 3D
calculations the model is suitable for, we have explored the stellar
orbital structure in the midplane of the Galaxy. We present the
standard analysis in the pattern rotating frame, and complement this
analysis with orbital information from the Galactic inertial
frame. Prograde and retrograde orbits are defined unambiguously in the
inertial frame, then labeled as such in the Poincar\'e diagrams of the
non-inertial frame. In this manner we found a sharp separatrix between
the two classes of orbits. Chaos is restricted to the prograde orbits,
and its onset occurs for the higher spiral perturbation considered
plausible in our Galaxy. An unrealistically high spiral perturbation
tends to destroy the separatrix and make chaos pervasive. This may be
relevant in other spiral galaxies.

\end{abstract}


\vfill\eject
\section{INTRODUCTION}

Modeling of spiral galaxies with sophisticated computational
techniques has become the usual way to study systems of this
nature. One of the important structures, which is in fact the one that
gives the name to this type of galaxies, is the spiral pattern. In the
spiral density wave theory (Lin \& Shu 1964), the spiral structure of
galaxies was modeled as a periodic perturbation term to the
axisymmetric potential in the disk plane. This is known as the {\it
tight-winding} or WKB approximation (e.g., Binney \& Tremaine 1994) for small
pitch angles. In the case of a two-armed spiral pattern it gives a
potential in the galactic plane of the form:

\be \Phi_{s} (R,\varphi) =  f(R)cos [ 2\varphi+g(R) ]; \label{eq_1}
\en

The function $f(R)$ is the amplitude of the perturbation, $g(R)$ provides
the geometry of the spiral pattern, and $R$, $\varphi$ are cylindrical
coordinates in the non-inertial reference frame of the arms (rotating
with a given angular velocity).

All studies of spiral galaxies we know of, even in cases of large pitch angle
in the spiral pattern, have used a spiral potential of the form in
Eq. (1), e.g. Contopoulos \& Grosb{\o}l (1986, 1988; hereafter C\&G86
and C\&G88); Patsis, Contopoulos, \& Grosb{\o}l (1991, hereafter PC\&G),
and in particular in our Galaxy the models of Amaral \& L\'epine
(1997, hereafter A\&L) and L\'epine, Mishurov, \& Dedikov (2001).
Self-consistency of the proposed spiral pattern has been analyzed by
C\&G86, C\&G88, PC\&G, and A\&L.

The dependence of the spiral potential on $z$ (perpendicular distance
to the galactic plane) has been accounted for by Patsis \& Grosb{\o}l (1996)
as a  $sech^2(z/z_s )$ factor of a function of the form in Eq. (1),
with $z_s$ a scaleheight. Martos \& Cox (1998), in numerical MHD
simulations, considered an exponential z-factor of an approximate
local spiral potential in the galactic plane.

In barred galaxies the approach is analogous to that given above for
spiral galaxies: the usual approximation for the potential in the
galactic plane due to the bar is a function of the form $\Phi_{b}
(R,\varphi) = f(R)cos ( 2\varphi)$. Instead of taking an ad hoc
dependence on the z coordinate, an alternative way to consider the
extension to a 3D-bar potential is to begin directly with a 3D mass
distribution representing the bar. This method has been considered by
Athanassoula et al. (1983) and Pfenniger (1984). From a comparison on
the galactic plane between their 3D-bar potential and a potential of
the form $\Phi_{b} (R,\varphi) = f(R)cos ( 2\varphi)$, Athanassoula et
al. (1983) found important differences in the corresponding force
fields. However, the consequences of this result were not pursued.

In this paper, rather than using a simple ad hoc model for a 3D spiral
perturbation, we consider a procedure whose essence is exactly the
same as the modeling of a barred galaxy made by Athanassoula \etal
(1983): instead of using a spiral potential of the form given by
Eq. (1), we propose a 3D mass distribution for the spiral arms and
derive their gravitational potential and force fields from previously
known results in potential theory. Grand design galaxies with a very
prominent spiral structure in red light suggest to us that such
structure should be considered an important galactic component and are
worthy of a modeling effort beyond a simple perturbing term. This
approach amounts to little more than admitting the possibility that
there is no simple formula that fits the spiral perturbation at all
$R$.

In our model we use Schmidt's (1956) analytical expression for the
potential of an inhomogeneous oblate spheroid and model the spiral
mass distribution as a series of such components settled along a
spiral locus. The overlapping of spheroids allows a smooth
distribution, resulting in a continuous function for the gravitational
force. The basic parameters of the excess density distribution
contributing to the spiral perturbation include a description of the
spiral locus, the dimensions and density law of the spheroids, the
central density in the spheroids as a function of galactocentric
distance, the total mass of the spiral arms, and the angular velocity
of the spiral pattern.

Our aim in this work is to make a preliminary study of stellar orbits
in the Galactic plane $z$ = 0 in a potential resulting from the
superposition of our 3D-spiral mass distribution and the axisymmetric
Galactic mass distribution considered by Allen \& Santill\'an (1991,
hereafter A\&S). Also, we compare the potential and force fields
produced by the 3D-spiral mass distribution with a tight-winding
approximation in Eq. (1). The resulting differences may have important
consequences on the stellar and gaseous dynamical behavior in a
potential of this type. An expected difference in the force
field is the effect of the self-gravitation of the mass of the spiral
arms, which is not accounted for in a potential like that of Eq. (1).

Detailed orbital studies have been made in barred and spiral galaxies
(e.g. Contopoulos 1983, Athanassoula et al. 1983, Pfenniger 1984,
Teuben \& Sanders 1985). In this work our analysis of stellar motion
in the Galactic plane, under the proposed Galactic potential, follows
the usual technique of Poincar\'e diagrams.  However, we propose an
alternative interpretation of Poincar\'e diagrams which has not been
previously considered. This interpretation is based on defining the
orbital sense of motion (prograde or retrograde) in the Galactic
inertial reference frame, joined to the usual definition in the
non-inertial reference frame (e.g. Athanassoula et al.  1983) in which
the spiral arms (and/or a bar) are at rest. This leads to Poincar\'e
diagrams (meaningful only in the non-inertial reference frame)
revealing two sharply separated regions: one corresponding to prograde
orbits and the other to retrograde orbits.  Our orbital analysis
emphasizes the properties of the Galactic spiral arms for which some
orbits may show stochastic behavior. These properties and the
resulting stellar behavior should be applicable to similar types of
galaxies.

The structure of this paper is the following: in Section 2 we present
our Galactic model for the 3D spiral arms, with a discussion of the
required parameters. In Section 3 we give the preliminary
self-consistency tests that we have made of the proposed spiral arms,
and establish a line of attack that must be followed to improve the
model. In section 4 we make a comparison between the potential and
force fields given by our model and those given by a tight-winding
approximation. We show the importance of the self-gravitation of the
spiral arms. In Section 5 we present an orbital analysis on the
Galactic plane for differing spiral arms properties including the
total mass in the spiral arms, the number of arms, and the angular
velocity of the spiral pattern. In Section 5.1 we clarify the
distinction between prograde and retrograde motion and the importance
of the frame of reference to establish the essential difference
between the two classes of orbits in Poincar\'e diagrams through the
zero angular momentum separatrix, a concept we introduce in this
section. We show here that our definition provides a direct connection
between sense of orbital motion and chaotic motion. In the same
subsection, Poincar\'e diagrams for a number of families labeled by
their Jacobi integral $E_J$ are shown.  An estimation of the required
strength of the spiral perturbation for which the nonlinear effects
are important is given, and we discuss the range of parameters
explored and those we deem plausible for our Galaxy. In Section 5.2 we
investigate the onset of chaos using Lyapunov exponents, and the
comparison of resonances for prograde and retrograde motion. In
Section 6 we discuss our results and give some conclusions, including
the possible response of the interstellar gas to the Galactic
potential.

\section{THE MODEL}
We use a Galactic model consisting of two mass distributions: the A\&S
axisymmetric model, and a 3D spiral model given by a superposition of
oblate inhomogeneous spheroids along a given locus. The A\&S Galactic
model assembles a bulge and a flattened disk proposed by Miyamoto \&
Nagai (1975), with a massive spherical halo extending to a radius of
100 kpc. The model is mathematically simple, with closed expressions
for the potential and continuous derivatives, which makes it
particularly suitable for numerical work. The model satisfies quite
well observational constraints such as the Galactic rotation curve and
the perpendicular force at the solar circle. The main adopted
parameters are $R_0 = 8.5$ kpc as the Sun's galactocentric distance,
and $V_0(R_0) = 220$ km s$^{-1}$ as the circular velocity at the Sun's
position. The total mass is $9\times 10^{11} M_{\odot}$, and the local
escape velocity is 536 km s$^{-1}$. The local total mass density is
$\rho_0 = 0.15 M_\odot$ pc$^{-3}$. The resulting values for Oort's
constants are $A = 12.95$ km s$^{-1}$ kpc$^{-1}$ and $B = -12.93$ km
s$^{-1}$ kpc$^{-1}$.
 
\begin{figure}
\plotone{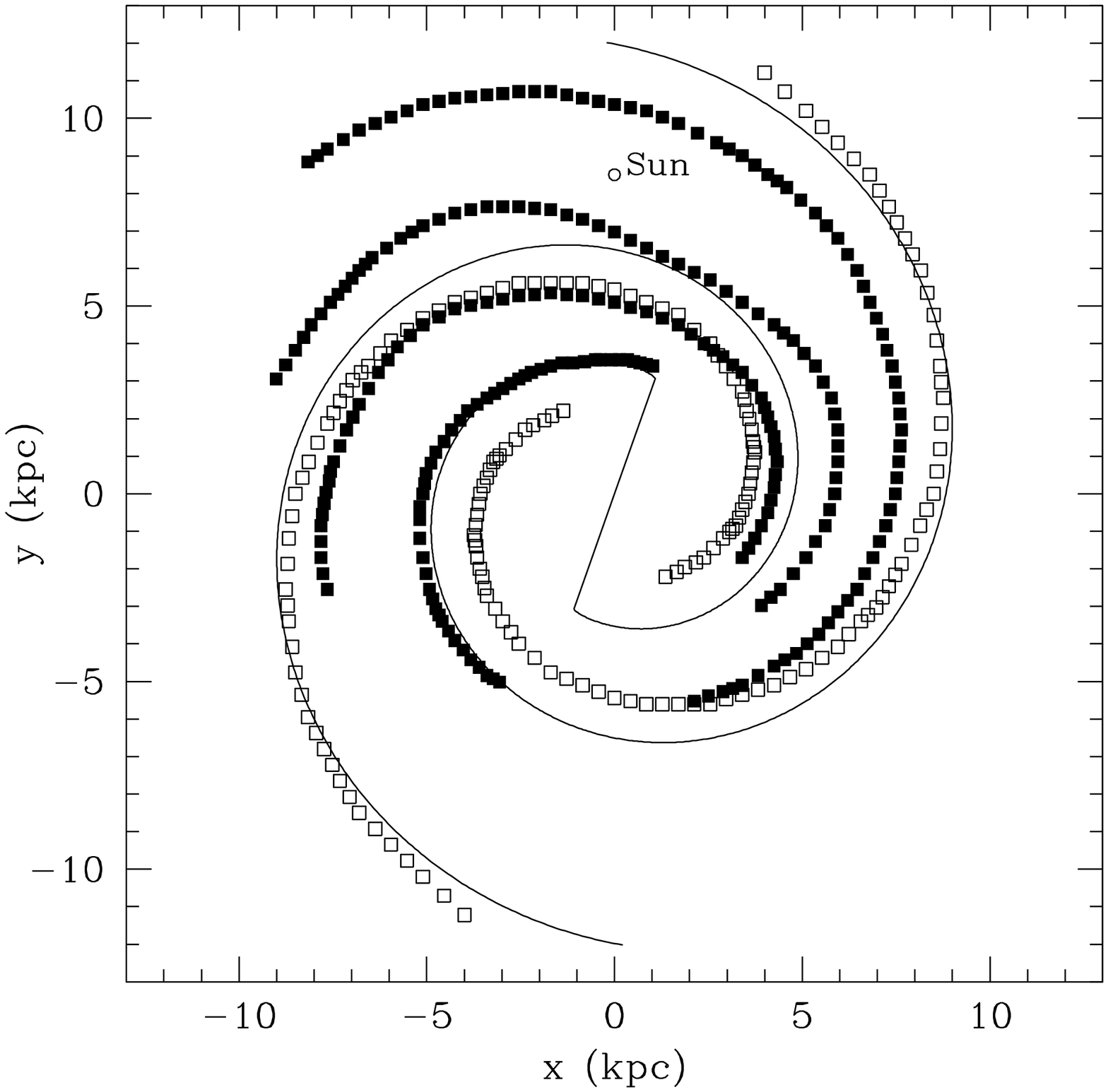}                      
\caption {Spiral arms in our Galaxy (Fig. 2 of Drimmel 2000). Optical
arms (black squares) and K-band arms (open squares). Continuous line:
first of three spiral loci considered to model the spiral arms.}
\label{locus1}
\end{figure}

As a first step to model the spiral mass distribution, we need the
spiral locus. The optical spiral structure in our Galaxy has been
studied by means of luminous HII regions (e.g, Georgelin \& Georgelin
1976, Caswell \& Haynes 1987). In the solar neighborhood the local
inclination (pitch angle) of this spiral pattern has been inferred
from the direction of the magnetic field lines (Heiles 1996), assumed
to be aligned with the spiral arms (see reviews by Beck 1993, and
Heiles 1995, on magnetic fields in spiral galaxies). In a recent
study, Drimmel (2000) presents evidence for a two-armed spiral in our
Galaxy as observed in the K band, which is associated with a
non-axisymmetric component in the old stellar population. Figure
\ref{locus1} reproduces Fig. 2 of Drimmel (2000); the black squares
trace the position of the four optical arms, and the open squares
represent his 15.5$^\circ$ pitch-angle fit for the two arms in the K
band. The continuous line shows the first of three spiral loci we
considered to model the spiral arms. These loci are obtained with a
function $g(R)$ (see Eq. 1) of the form given by Roberts, Huntley \&
van Albada (1979),

\be g(R) =-(\frac{2}{N\ {\rm tan}\ i_p})\ {\rm ln}\ [1+(R/R_s)^N] \label{eq_2}
\en

\noindent with $i_p$ the pitch angle at $R$ $\rightarrow \infty$. In
Figure \ref{locus1} we take N=100, thus making the arms start at a
distance $R_s$ and at right angles to a line passing through the
Galactic center, $i_p$=11$^\circ$, $R_s$=3.3 kpc, and we consider an
orientation such that the two arms start on a line making an angle of
20$^\circ$ with the Sun-Galactic center line (this is the approximate
direction of the Galactic bar, e.g. Freudenreich 1998). This first
two-armed spiral locus approximates the position of both optical and
K-band arms.

\begin{figure}
\plotone{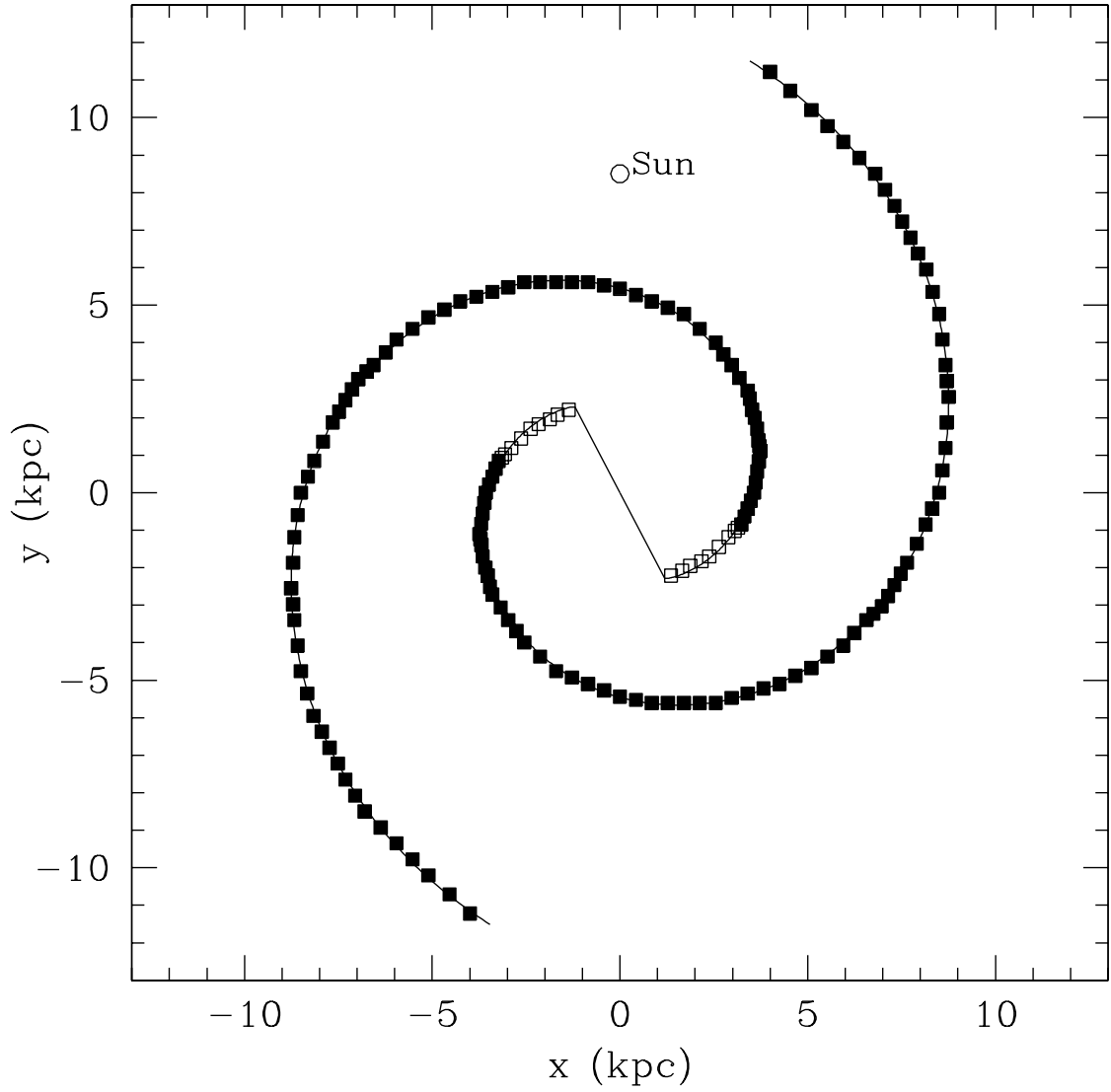}                      
\caption {Only the K-band spiral arms. The continuous line is our 
second spiral locus. We consider the region $R \geq 3.3$ kpc, traced
by the black squares.}
\label{locus2}
\end{figure}

As a second spiral locus, we take the two-armed, K-band locus
itself. In Figure \ref{locus2} the black and open squares give the
K-band arms in Figure \ref{locus1}. The continuous line is obtained
with the function $g(R)$ in Eq. (2), taking N = 100, $i_p$ =
15.5$^\circ$, and $R_s$ = 2.6 kpc. We consider the effective starting
distance of the spiral arms in this second locus as the distance 3.3
kpc taken in the first locus; the inner black squares in Figure
\ref{locus2} mark this position.

\begin{figure}
\plotone{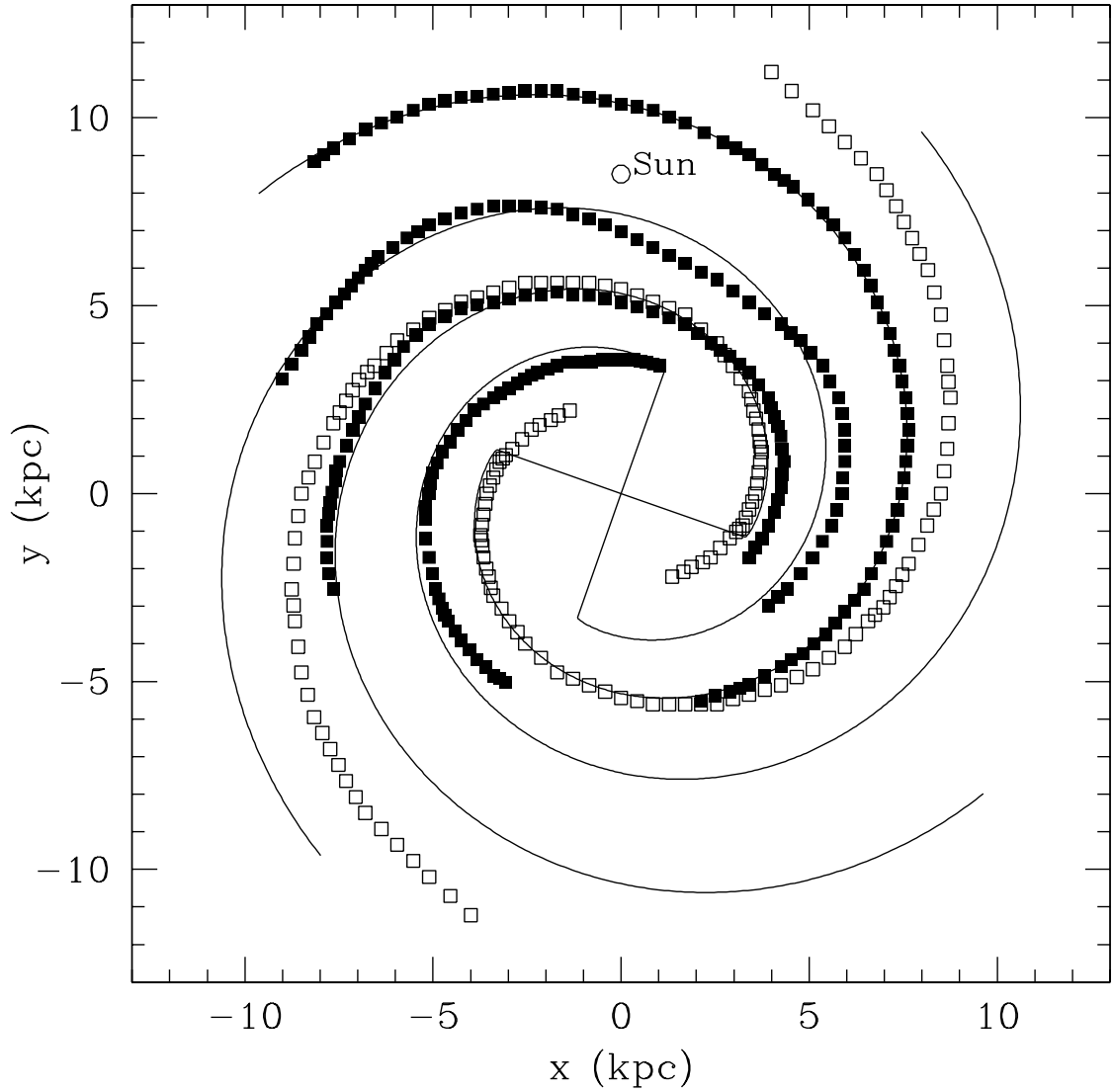}                      
\caption {Spiral arms in our Galaxy. Optical arms (black squares) and 
K-band arms (open squares). Continuous line: is our fit to a
four-armed locus. The third locus we considered is composed by the two
locus joined, our fit to K-band arms (shown in the previous figure)
and the four-armed locus presented in this Figure.}
\label{locus3}
\end{figure}

An important spiral locus is the one given by the four optical arms.
Vall\'ee (2002) shows in his Fig. 2 his fit to these optical arms,
taking a pitch angle of 12$^\circ$ and $R_0$ = 7.2 kpc for the Sun's
galactocentric distance. Figure \ref{locus3} shows our fit with N =
100, $i_p$ = 12$^\circ$, and $R_s$ = 3.54 kpc in Eq. (2), and $R_0$ =
8.5 kpc. Drimmel (2000) has suggested that the four optical arms in
our Galaxy trace the response of the gas to the two-armed, K-band,
stellar spiral arms in Figure \ref{locus2}. Thus, we take the third
spiral locus for the spiral arms in our Galaxy as the superposition of
the two-armed, K-band and four-armed, optical spiral loci. Galactic
models using a superposition of two- and four-armed spirals have been
proposed by A\&L and L\'epine, Mishurov, \& Dedikov (2001).

Our set of models for the spiral mass distribution consist of a
superposition of oblate inhomogeneous spheroids along each of the
three proposed spiral loci.  The minor axis of each spheroid is
perpendicular to the Galactic plane.  Each spheroid has a similar mass
distribution, i.e., surfaces of equal density are concentric spheroids
of constant semi-axis ratio.  We consider a linear density law
$\rho(a) = p_0 + p_1a$ in the spheroids, with $a$ the major semi-axis
of a similar surface, and the coefficients $p_0$, $p_1$ being
functions of the galactocentric distance of the spheroid's
center. Schmidt (1956) has given the expressions for the potential and
force fields for a spheroid with this density law.

With respect to the dimensions of the spheroids, Kennicutt \& Hodge
(1982) have analyzed a sample of spiral galaxies; from their Fig. 4,
the average width of the spiral arms is around 1 kpc. Thus,
considering the linear fall in the density within the spheroids, and
taking the vertical extension of the spiral arms as the mean
scaleheight (the vertical structure of the arms is discussed in Martos
\& Cox 1998), in most models we take the minor ($c_0$) and major
($a_0$) semi-axes of the oblate spheroids as 0.5 kpc and 1.0 kpc,
respectively, with a separation of spheroid centers along the spiral
locus of 0.5 kpc. We found no significant change in our results if we
decrease this separation (thus increasing the smoothness of the spiral
mass distribution). Each spheroid has zero density at its boundary;
thus the coefficients $p_0$, $p_1$ in a given spheroid satisfy
$p_0$(R) = -$a_0$$p_1$(R), with R the galactocentric distance of the
spheroid's center. The function $p_0$(R) is discussed below.

The superposition of spheroids begins at the distance $R_i$ = 3.3 kpc
in the first and second spiral locus given above, and at $R_i$ = 3.5
kpc in the third locus. The spiral arms are truncated, i.e., the
superposition of spheroids ends, at a distance $R_f$. The analyses of
C\&G86, C\&G88, and PC\&G establish that for strong spirals nonlinear
effects make self-consistent spirals terminate at the 4/1 resonance,
in contrast with weak spirals in which linear theory predicts they can
extend up to or beyond the corotation resonance (e.g., Vauterin
\& Dejonghe 1996; Kikuchi, Korchagin, \& Miyama 1997). According to
PC\&G, strong spirals are those in which the force produced by the
spiral perturbation is greater than 6$\%$ of the background force. We will
consider models around and above this limit, and hence the distance
$R_f$ should be taken in accordance with these results. However, our
main criterion to set the value of $R_f$ is the maximum radial extent
of the observed spiral arms shown in Figure \ref{locus1}.  In all
models we take the value $R_f$ = 12 kpc; in Figures
\ref{locus1}-\ref{locus3} the continuous lines end at this distance.

In our models we consider two functions $p_0(R)$ for the
central density in the spheroids, defined in the interval $R_i \leq R
\leq R_f$ : (1) a linear fall to zero $p_0(R)
=p_{01}(R_f-R)/(R_f-R_i)$, and (2) an exponential fall $p_0(R) =
p_{02}e^{-(R-R_i)/R_L}$, with $R_L = 2.5$ kpc the approximate radial
scalelength of the near-infrared Galactic disk (Freudenreich 1998).
We compare the linear fall to the exponential one; the latter being
the form generally employed in studies of spiral galaxies (e.g.,
C\&G86, C\&G88, PC\&G, A\&L, Patsis \& Grosb{\o}l 1996, Englmaier \&
Gerhard 1999).

The values of the coefficients $p_{01}$, $p_{02}$ are

\be p_{01} = \frac{3 M_s (R_f-R_i)}{2 \pi\ a_0^2\ c_0 \sum_{j=1}^{N_t} 
(R_f-R_j)}     \label{}
\en

\be p_{02} = \frac{3 M_s}{2 \pi\ a_0^2\ c_0 \sum_{j=1}^{N_t} e^{-(R_j-R_i)/R_L}} 
    \label{eq_3}
\en

\noindent with $M_s$ the total mass in the spiral arms, $N_t$ the
total number of spheroids in each arm, and $R_j$ the galactocentric
distance of spheroids' centers. The sums are only over one arm.  In
a model with four arms we multiply the terms on the right by a
factor 1/2.
 
With the above expressions of $p_0$(R), and $p_1$(R) =
-$p_0$(R)/$a_0$, Schmidt's (1956) equations give the potential and
force produced by a spheroid at any point in space; the corresponding
total potential and force are obtained by summing over all spheroids in
all the arms.

The ratio of the total mass in the spiral arms to the mass of the disk
(M$_D$ = $8.56\times 10^{10}$ M$_\odot$) in the A\&S Galactic model is
taken to be $M_S/M_D$ = 0.0175, 0.03, and 0.05. In the next section
some properties of the models based on these values are analyzed.

A final parameter in the models is the angular velocity of the 
spiral arms, $\Omega_p$, which we assume to be rigidly rotating. 
The Galactic model of A\&L
favors the value ${\Omega}_{p}$ = 20~km~s$^{-1}$~kpc$^{-1}$, but the
hydrodynamical calculations of Englmaier \& Gerhard (1999) and Fux
(1999), giving the gaseous response in the Galactic disk to a Galactic
barred potential, suggest that $\Omega_p$ might be as large as
60~km~s$^{-1}$~kpc$^{-1}$. However, Englmaier \& Gerhard (1999) point
out that the Galactic spiral arms and the Galactic bar might not have
the same pattern speed. Thus, in our models we take a clockwise
rotation, and consider ${\Omega}_{p}$ = 20, 60~km~s$^{-1}$~kpc$^{-1}$
as two possible values for the pattern speed of the spiral arms. For
comparison, we have also included computations with a plausible
smaller value of ${\Omega}_{p}$ =15~km~s$^{-1}$~kpc$^{-1}$ (see
discussions in Martos \& Cox 1998; Gordon 1978; Palous \etal 1977; Lin,
Yuan, \& Shu 1969). Figure \ref{resonancias1} gives some resonance
curves in the A\&S Galactic model. In the case of a
two-armed spiral pattern, and with ${\Omega}_{p}$ =
15~km~s$^{-1}$~kpc$^{-1}$, the inner Lindblad resonance is at 3.5 kpc,
corotation at 14.3 kpc, the external Lindblad resonance at 23.2 kpc,
and the 4/1 resonance at 9.5 kpc. The corresponding values with
${\Omega}_{p}$ = 20, 60~km~s$^{-1}$~kpc$^{-1}$ are 2.8, 10.9, 17.7,
and 7 kpc; 1.36, 3.38, 6.28, and 2.21 kpc.

\section{SELF-CONSISTENCY ANALYSIS}

The self-consistency of a stationary spiral pattern, as the one here
proposed, must be addressed. PC\&G constructed self-consistent models
for twelve normal spiral galaxies. The sample included Sa, Sb, and Sc
types. Their conclusion is that for the Sb and Sc galaxies, the best
self-consistent model is a nonlinear one in which the 4/1 resonance
determines the distance beyond which the response density does not
enhance the spiral, that is, the extent of the spiral pattern. Figure
15 of PC\&G shows an approximate correlation in self-consistent models
between the pitch angle of the spiral arms, $i_p$, and the relative
radial force perturbation (absolute value of the ratio of radial
forces produced by the spiral arms and the background, both evaluated
at each point). According to that figure, our Galaxy, with $i_p$
$\sim$ 15$^\circ$, would require for a self-consistent model a
relative force perturbation between 5 and 10$\%$. In our models the
ratio $M_S/M_D$ was chosen within limits suggested by PC\&G result. We
take $M_S/M_D$ = 0.0175, 0.03, and 0.05, which imply a peak relative
force perturbation of approximately 6, 10, and 15$\%$, and average
values over $R$ \footnote[1]{Notice that the force is a sensitive
function of $R$. See Section 6 for a discussion on the consequences of
that fact} of approximately 3, 6, and 10$\%$, respectively.

Figure \ref{fuerzas} plots the radial force produced by the spiral
arms in our model and the corresponding relative force perturbation,
as functions of galactocentric distance R. In this figure the mass
ratio is $M_S/M_D$ = 0.0175 and $p_0$(R) has the exponential fall with
a scalelength of 2.5 kpc; similar results are found with the linear
fall of $p_0$(R).  In the left frames of the figure the model has the
spiral locus in Figure \ref{locus1}; in the right frames the spiral
locus is that of Figure \ref{locus2}. In each case, the radial force
(scaled by the absolute value of the force given by the A\&S model at
the Solar position) and the relative force perturbation are given
along two radial lines (we call these two lines the $x'$ and $y'$
axes, respectively; see Figure \ref{equipotenciales}): the line
passing through the starting points of the spiral arms (continuous
curves in the figure), and the line at right angles (dotted curves in
the figure). The radial force, in the upper frames, shows the sign
changes along the two chosen radial directions. The relative force
perturbation is shown in the lower frames. Similar figures are
obtained in the cases $M_S/M_D$ = 0.03 and 0.05, showing the
corresponding peak values quoted above for the relative force
perturbation.

We made a preliminary study of self-consistency in the models of
Figure \ref{fuerzas}, using the two values of ${\Omega}_{p}$,
15~km~s$^{-1}$~kpc$^{-1}$ and 20~km~s$^{-1}$~kpc$^{-1}$. We followed
the method of C\&G86 to obtain the density response to the given spiral
perturbation. This method assumes that the stars with orbits trapped
around an unperturbed circular orbit, and with the sense of rotation
of the spiral perturbation, are also trapped around the corresponding
central periodic orbit in the presence of the perturbation. Thus, we
computed a series of central periodic orbits, and found the density
response along their extension, using the conservation of mass flux
between any two successive orbits. The initial circular orbits were
taken with a separation of 0.25 kpc. For more details see C\&G86.

We found the position of the maxima density response along each
periodic orbit, and thus the positions of the response maxima on the
Galactic plane are known. These positions are to be compared with the
center of the assumed spiral arms, i.e. the spiral locus.

Figure \ref{periodicas} illustrates some results. This figure shows
the positions of the response maxima (black squares), along with the
spiral arms (open squares), and the periodic orbits used in the
method. Cases (a), (b), (c) have the spiral locus of Figure
\ref{locus2}, and case (d) the spiral locus of Figure
\ref{locus1}. The value of ${\Omega}_{p}$ is 15~km~s$^{-1}$~kpc$^{-1}$
in case (a), and 20~km~s$^{-1}$~kpc$^{-1}$ in cases (b), (c), (d). The
mass ratio is $M_S/M_D$ = 0.0175 in cases (a), (b), (d), and $M_S/M_D$
= 0.00875 in case (c).

Figure \ref{periodicas} shows that mostly the response maxima lag
behind the spiral arms, as the galactocentric distance increases. This
behavior has already been discussed by C\&G88 and PC\&G: the response
maxima may lag behind or ahead the spiral arms, depending on its
radial scalelength and strength, among other parameters. PC\&G give a
nonlinear self-consistent model for the spiral galaxy NGC 1087, in
which the response maxima lag behind, and claim that this is
consistent with observations. C\&G88 show that if they consider a
dispersion of velocities around the central periodic orbits, the
displacement between the response maxima and the spiral arms
diminishes, obtaining a better self-consistency.

Once we found the positions of the response maxima, with the density
response along each central periodic orbit we computed the average
density response ${\rho}_{resp}$ around each one of these positions,
taking a circular vicinity of radius equal to the semi-axis $a_0$ of
the spheroids in the model. We then compared ${\rho}_{resp}$ with the
imposed density, i.e. the one proposed by the model. This imposed
density, ${\rho}_{mod}$, is the sum of the A\&S disk density on the
Galactic plane and the central density of the spiral
arms. ${\rho}_{mod}$ is computed $along$ the arms. The densities
${\rho}_{resp}$ and ${\rho}_{mod}$ are compared at a same
galactocentric distance $R$, but the corresponding positions on the
Galactic plane may differ in azimuth, as shown in
Fig. \ref{periodicas}.

Following C\&G88 and A\&L, we computed the ratio
${\rho}_{resp}$/${\rho}_{mod}$ to analyze the self-consistency of the
assumed spiral perturbation. This ratio should be close to 1. Figure
\ref{contraste} shows the value of this ratio for each case in
Fig. \ref{periodicas}. Figure \ref{contraste}(a), with ${\Omega}_{p}$
= 15~km~s$^{-1}$~kpc$^{-1}$, shows a high density response in the
inner region where the spiral arms begin. This behavior has been
discussed by C\&G86 and C\&G88. However, Figures
\ref{contraste}(b),(c),(d), with ${\Omega}_{p}$ =
20~km~s$^{-1}$~kpc$^{-1}$, show a lower response in the inner region,
making the ratio ${\rho}_{resp}$/${\rho}_{mod}$ be closer to 1. Thus,
in this preliminary analysis, we favor ${\Omega}_{p}$ =
20~km~s$^{-1}$~kpc$^{-1}$. In contrast, A\&L found (for their model)
that the density response could not favor a specific value of
${\Omega}_{p}$.

Case (c) in Fig. \ref{contraste} shows a better self-consistency than
case (b), which has twice the mass in the spiral arms. In both cases the
density response differs strongly from the imposed density in the region
around the distance (7 kpc) of the corresponding 4/1 resonance. Case (d),
with the spiral locus of Fig. \ref{locus1}, appears to give an
acceptable density response, even in the region of the 4/1 resonance.

The approximately self-consistent four models (a)-(d) of
Figs. \ref{periodicas} and \ref{contraste} have mass in the spiral
arms seemingly in the lower limit of the interval (from Fig. 15 of
PC\&G) which an Sb galaxy like ours needs to sustain a nonlinear,
self-consistent spiral perturbation. The analysis for self-consistency
is also needed around the upper limit of the mass ratio $M_S/M_D$. As
the method of C\&G86 gives only approximate results (C\&G88), this
analysis and the preliminary results given above need to be
reconsidered using the suggested improvements for self-consistency
found by C\&G88: it may be necessary to account for a population of
orbits around the periodic orbits, with an appropriate velocity
dispersion, and perhaps also the inclusion of four-armed spirals
(A\&L, L\'epine, Mishurov, \& Dedikov 2001). In a study currently
underway, we are exploring the self-consistency of our models
considering these components, with the mass ratio $M_S/M_D$ in the
suggested interval, with other density laws in the oblate spheroids,
and taking the dimensions $a_0$, $c_0$ as functions of galactocentric
distance. A 3D orbital analysis as the one made by Patsis \&
Grosb{\o}l (1996) would be also relevant in this procedure.

\section{A COMPARISON WITH THE TIGHT-WINDING APPROXIMATION}

In the tight-winding approximation for the spiral arms (e.g., Binney \&
Tremaine 1994), the potential at a given point is determined by the
properties of the spiral arms in a small vicinity around the point.
This approximation is given by Eq. (1) in the case of a two-armed
spiral pattern. In our model, the potential at any point in space is
obtained by summing the contributions of every element of mass along
the spiral arms. Thus, a comparison of our model with the
tight-winding model has some interest.

In Figure \ref{TW1} we give the potential and radial force, scaled by
the absolute value of the potential and radial force of the A\&S model
at the Solar position, of a model (continuous line) with the spiral
locus in Figure \ref{locus2}, a mass ratio $M_S/M_D$ = 0.0175, and an
exponential fall of the central density in spheroids, $p_0$(R), with a
radial scalelength of 2.5 kpc. We plot the potential and radial force
of the spiral arms along three radial lines: the positive $x'$ axis
(upper frames), and the lines at 60$^\circ$ (middle frames) and
120$^\circ$ (lower frames) from the $x'$ axis (in the direction toward
the $y'$ axis). The dotted line shows the corresponding potential and
radial force of a tight-winding model, i.e., Eq. (1), with the same
spiral locus as in our model (Fig. \ref{locus1}), and with an
amplitude f(R) (Eq. (1)) of the form considered by C\&G86: f(R) =
-AR$e^{-{\epsilon}_{s}R}$. In Figure \ref{TW1} we take A =
450~km$^{2}$~s$^{-2}$~kpc$^{-1}$ and ${\epsilon}_{s}$ = 1/2.5
kpc$^{-1}$ (that is, the same radial scalelength as in our model). The
high 15.5$^\circ$ pitch angle of the spiral locus is not suitable for
a rigorous comparison, but we see that our model cannot be
well-approximated by a tight-winding model. Notice, for instance, in
the upper right frame of Figure
\ref{TW1} the effect of the mass in the spiral arms inner to $\sim$ 8
kpc: the attraction of the whole spiral pattern requires that the
point at which the radial force changes from negative to positive
needs to be closer to the spiral arm around 8 kpc than in the
tight-winding model. This accumulated negative radial force shifts the
net force toward negative values in the outer regions.

In Figure \ref{TW2} we compare the radial forces along the positive
$x'$ axis, for models with a 3$^\circ$ pitch-angle, two-armed, spiral
locus starting at 3.3 kpc. The continuous line gives our model with
$a_0$ = 0.1 kpc, $c_0$ = 0.05 kpc, separation of spheroids' centers of
0.05 kpc, a low mass ratio $M_S/M_D$ = 0.001, and an exponential fall
of $p_0$(R) with a radial scalelength of 2.5 kpc. The dashed line
results from a tight-winding model with A =
9~km$^{2}$~s$^{-2}$~kpc$^{-1}$ and ${\epsilon}_{s}$ = 1/2.5
kpc$^{-1}$. Both models are similar in the inner region, but as the
galactocentric distance increases the effect mentioned above begins to
be important.  Even in this low-mass case, the attraction of all the
mass in the spiral arms makes the radial force in the outer regions
asymmetric around zero; in fact, at large distances this force (per
unit mass) is -$GM_{S}/R^2$. In Section 6 we discuss briefly some
consequences of these results, which must have important consequences
to the gas dynamics.

\section{ORBITAL ANALYSIS}

As an application of our model, we have made a brief study of stellar
orbits in the Galactic plane. Poincar\'e diagrams are presented, and
discussed from the perspective of the sense of orbital motion defined
in the Galactic inertial frame and its connection to the onset of
stochastic motion. To investigate further the nature of chaotic motion
apparent in Poincar\'e diagrams, we utilized additionally Lyapunov
exponents (Wolf 1984).

\subsection{Poincar\'e Diagrams and the Separatrix of Zero Angular
Momentum}

The orbital analysis is made in the non-inertial reference frame
attached to the spiral pattern, labeled as the primed system of
Cartesian coordinates $(x',y',z')$. As defined in Section 3, the $x'$
axis is taken as the line passing through the inner starting points of
the spiral arms; the $z'$ axis is perpendicular to the Galactic plane,
with its positive sense toward the north Galactic pole, and the $y'$
axis is such that the $(x',y',z')$ axes form a right-hand system. The
angular velocity of the spiral arms, ${\Omega}_{p}$, points in the
negative direction of the $z'$ axis, i.e., a clockwise rotation.

In the Galactic plane the effective potential in the non-inertial
frame is given by

\be {\Phi}_{eff} (x',y') = {\Phi}_{AS} (x',y') + {\Phi}_{s} (x',y') -
(1/2){\Omega}_{p}^2 (x'^ 2 +y'^2), \label{eq_5} \en

\noindent with ${\Phi}_{AS}$ the A\&S potential and ${\Phi}_{s}$ the
potential due to the spiral arms.

Figure \ref{equipotenciales} shows some equipotential curves
${\Phi}_{eff}$ = const. for the model with $M_S/M_D = 0.0175$,
${\Omega}_{p}$ = 20 km s$^{-1}$~kpc$^{-1}$, exponential fall of
$p_0$(R), and the spiral locus in Figure \ref{locus2} (i.e.,
$i_p$=15.5$^\circ$). Both figures  \ref{locus2} and
\ref{equipotenciales} have the same orientation of the spiral
pattern. The inertial $x,y$ and non-inertial $x',y'$ axes are shown in
Figure \ref{equipotenciales}. Each square traces the center of an
oblate spheroid, and the islands in the equipotential curves appear at
the corotation distance (10.9 kpc in this case).

A known integral of stellar motion in the non-inertial system is
Jacobi's expression $E_J$ = $(1/2){\rm{\boldmath v'}}^2 +
{\Phi}_{eff}$, with ${\bf v}'$ the velocity in this system. Then the
equipotential curves are curves of zero velocity for corresponding
values of $E_J$. Figure \ref{Jvsx} plots the value $E_J$ =
${\Phi}_{eff}$ on the positive $x'$ axis, for the model in
Figure \ref{equipotenciales}.

Poincar\'e diagrams were constructed following the usual procedure. We
found the crossing points with the $x'$ axis of orbits with a given
value of $E_J$, and made Poincar\'e diagrams $x'$ vs $v_x'$ for the
crossing points having $v_y' > 0$ . All the crossing points with $v_y'
< 0$ were incorporated in the $v_y' > 0$ diagrams taking $x'
\rightarrow -x'$ , $v_x' \rightarrow -v_x'$. We studied several models
with a two-armed spiral pattern, taking combinations of ${\Omega}_{p}$
(15 or 20 km s$^{-1}$ kpc$^{-1}$), $i_p$ (11$^\circ$ or 15.5$^\circ$),
$M_S/M_D$ (0.0175, 0.03, or 0.05), and the function $p_0$ (linear or
exponential). Also, we studied models with the $six$ spiral arms in
Figure \ref{locus3}, taking combinations of ${\Omega}_{p}$ = 20, 60 km
s$^{-1}$ kpc$^{-1}$, $M_S/M_D$ = 0.0175, 0.05, and an exponential
function $p_0$. In all cases the orbits were computed with a
Bulirsch-Stoer algorithm (Press \etal 1992), with a mean maximum error
$|(E_{J_{final}}- E_{J_{initial}})/E_{J_{initial}}|$ of order
$10^{-12}$, in runs with elapsed physical times of $10^{9}$ to
$10^{11}$ years.

In this subsection we present Poincar\'e diagrams for models with the
lower mass ratio $M_S/M_D$ = 0.0175; results with $M_S/M_D$ = 0.05 are
presented in the next section.

Figure \ref{poincare9} shows Poincar\'e diagrams for the model of
Figures \ref{equipotenciales} and \ref{Jvsx}, i.e., ${\Omega}_{p}$ = 20
km s$^{-1}$ kpc$^{-1}$ and the spiral locus in Figure \ref{locus2}
($i_p$=15.5$^\circ$). Values of $E_J$ were selected in the interval
[-1.8, -1.2]$\times$~$10^5 $~km$^{2}$~s$^{-2}$. In this and following
figures each diagram was constructed with approximately fifty
orbits. The boundary of the permitted region begins to open at the
value $E_J$ of the island around the corotation distance (see Figure
\ref{equipotenciales}) to which the $x'$ axis is tangent. This value
of $E_J$ is the maximum of the curve in Figure \ref{Jvsx}.

Figure \ref{6brazosOMP2} gives four Poincar\'e diagrams for a model
with the six spiral arms in Figure \ref{locus3}, and again
${\Omega}_{p}$ = 20 km s$^{-1}$ kpc$^{-1}$. In this example the mass
ratio for the two, K-band, spiral arms is $M_S/M_D$ = 0.0175, and the
total mass in the four optical arms has this same ratio; i.e. the
total mass in the four optical arms is equal to the total mass in the
two K-band arms.

In the two Poincar\'e diagrams shown in Figures \ref{MSD0.05_J2200}
and \ref{MSD0.05_J2050}, we keep the same spiral pattern arms as in
Figure \ref{6brazosOMP2}, but take the high pattern speed
${\Omega}_{p}$ = 60 km s$^{-1}$ kpc$^{-1}$. Figure
\ref{equipotenciales6} shows some ${\Phi}_{eff}$ = const. curves for
this case. The black squares give the four optical arms, and the open
squares the two K-band arms. The $x'$ axis in this case of six arms
is the starting line of the two K-band arms. The islands in the
${\Phi}_{eff}$ = const. curves appear at a lower corotation distance
(c.f., Figure \ref{equipotenciales}). Figure \ref{Jvsx6} gives
$E_J$ = ${\Phi}_{eff}$ on the positive $x'$ axis.

The orbital structure of Poincar\'e diagrams shown in Figures
\ref{poincare9} to \ref{MSD0.05_J2050}, is the usual structure
obtained in studies of stellar orbits in spiral and barred galaxies
(e.g., Contopoulos 1983, Athanassoula et al. 1983, Teuben \& Sanders
1985). A dominant periodic orbit appears in the $x' > 0$ side of each
diagram, and in some cases (i.e., for a certain range in $E_J$) there
is also a dominant periodic orbit on the $x' < 0$ side. Rather than a
detailed analysis of the orbital structure, what we wish to emphasize
in these diagrams is the clear separation of two regions, each one
containing orbits with a definite sense of rotation, prograde or
retrograde, defined in the Galactic inertial frame. In our situation
the spiral pattern moves in the clockwise sense; so the usual
definition in the non-inertial frame (e.g., Athanassoula et al. 1983)
would call prograde orbits those orbits crossing the $x' < 0$ side,
and retrograde orbits those crossing the $x' > 0$ side. This
definition is ambiguous, because the azimuthal velocity in the
non-inertial frame may change sign along a given orbit. Thus, an
orbit may be both prograde and retrograde (Contopoulos 1983). On the
other hand, in the inertial frame, {\it and for the considered range
of strengths of the spiral perturbation}, orbits maintain the sign of
their azimuthal velocities, with the exception of orbits with angular
momentum close to zero (as computed in this frame).

We define the sense of orbital rotation in the inertial frame, as
follows: prograde if the azimuthal velocity is always of the same sign
as the angular velocity of the spiral pattern, $\Omega_p$; and retrograde
if the azimuthal velocity always has the opposite sign of $\Omega_p$.
With this definition, Poincar\'e diagrams show a sharp separation
between the regions of prograde and retrograde orbits by a ``curve''
that we call the separatrix of zero angular momentum, which
corresponds to orbits with nearly vanishing angular momentum in the
inertial frame. In Figures \ref{poincare9} to \ref{MSD0.05_J2050}, the
separatrix is shown with darker spots.  The orbits forming this
``curve'' would need to be computed over a longer time to fill it in the
diagrams where it appears to be discontinuous.

The definition of sense of orbital motion in the inertial frame
reveals that all orbits inside the region bounded by the separatrix
are retrograde.  Prograde orbits are outside this region. Prograde
orbits may have points on both the $x' > 0$ and $x' < 0$ sides of a
Poincar\'e diagram. Another way of saying this is that only prograde
orbits may change their sense of motion in the rotating frame. The
separatrix is the transition region between prograde and retrograde
orbits in the inertial reference frame.

In Figure \ref{MSD0.05_J2050} the prograde region has orbits with
appreciable chaotic motion. This behavior is discussed in more detail
in the next subsection.

We have stressed above that the definition of sense of orbital
rotation in the Galactic inertial frame is useful, as long as the
strength of the spiral perturbation is not too high. From our study,
the correlation chaos-prograde motion seems valid to our Galaxy. In
the next section we analyze the orbital structure in the case of the
higher mass ratio $M_S/M_D$ = 0.05 still applicable to our Galaxy in
our framework. We will see that the separatrix increases its width,
i.e., the number of orbits which are both prograde and retrograde in
the inertial frame increases. However, our definition still provides a
clear separation of prograde and retrograde orbits.  An interesting
situation in which this definition apparently loses its usefulness is
the case of lopsided galaxies considered by Noordermeer, Sparke, \&
Levine (2001). In this case it is expected that even in the inertial
frame there is a wide region of orbits that are both prograde and
retrograde.

Defining the sense of orbital motion in the inertial frame has a more
physical connection with the character of orbits affected by a spiral
perturbation. As we will see in the next section, the analysis of 
Poincar\'e diagrams based on this definition shows a connection
with the onset of chaotic motion.

\subsection{Exploring the Nature of Orbital Chaos}

Orbital chaos has been found in potentials including spiral or bar
perturbations (e.g., Contopoulos 1983, Athanassoula et al. 1983,
Pfenniger 1984, Teuben \& Sanders 1985, Fux 2001), and in other
non-axisymmetric potentials (e.g., Alvarellos 1996; Noordermeer,
Sparke, \& Levine 2001). In this subsection we present some results
related with the onset of chaotic stellar motion in our model, taking
the higher mass ratio $M_S/M_D$ = 0.05, which we consider applicable
in our Galaxy.  We find that as the mass ratio $M_S/M_D$ increases,
the onset of orbital chaos always occurs outside the region bounded by
the separatrix defined in the previous section, i.e., the prograde
region.  Furthermore, within the plausible range of $M_S/M_D$, chaos
is entirely confined to the prograde region.

In Figure \ref{MSD0.0175} we give a Poincar\'e diagram with $E_J =
-1630 \times 10^2 ~{\rm km}^2 {\rm s}^{-2}$, (previously shown in
Figure 12, with $M_S/M_D$ = 0.0175), and in Figure \ref{MSD0.05} a
diagram with the same value of $E_J$, but now with $M_S/M_D$ =
0.05. In both cases the pattern speed is ${\Omega}_{p}$ = 20 km
s$^{-1}$ kpc$^{-1}$. A comparison of the two figures shows that
increasing the strength of the spiral perturbation causes the
separatrix to increase in width, and also some chaotic motion begins
to appear on the $x' < 0$ side in Figure \ref{MSD0.05}, i.e., outside
the region bounded by the separatrix.

To investigate the orbital chaos which appears in Figure
\ref{MSD0.05}, we did a Lyapunov exponents analysis following Wolf
(1984), by calculating individual orbits in the diagram in Figure
\ref{MSD0.05}. The first Lyapunov exponent was calculated to classify
orbits as chaotic or non-chaotic by applying the usual criterion for
chaos, namely $\lambda > 0$ for chaotic motion (which means that two
orbits with very close initial conditions, will increase their
relative distance as $e^{\lambda t}$, with $t$ the time), and $\lambda
\leq 0$ for regular motion (where the relative distance will be
constant or decrease exponentially to zero). We found that the
exponent $\lambda$ in the prograde region, in the scattered-points
subregions, such as that marked with the number 1 in Figure
\ref{MSD0.05}, is always positive ($\ga .4$) for each pair of orbits
we tried, as expected for conservative chaos. On the other hand, this
exponent is less than zero for orbits in the regular regions of the
Poincar\'e diagram for both prograde and retrograde orbits (marked
with numbers 2 and 3, respectively). Thus, chaotic motion is entirely
confined to the prograde orbits (seen from the inertial frame) for
plausible parameters for the spiral arms in the Galaxy.

If we further increase the mass ratio $M_S/M_D$, we find that chaotic
motion is important for a range of values of $E_J$, and it spreads
from the prograde region toward the retrograde region; This behavior
was seen in previous orbital studies (e.g. Contopoulos 1983,
Athanassoula et al. 1983, Teuben \& Sanders 1985), although not linked
with the sense of orbital motion defined in the inertial frame.

Figure \ref{6brOM2J1630} shows the effect which produces the addition
of the four optical arms to the two K-band arms considered in Figure
\ref{MSD0.05}, the total mass in the optical arms being the same as
the mass in the K-band arms. There are structural similarities between
both diagrams (Figs. \ref{MSD0.05} and Fig. \ref{6brOM2J1630}), but the
main difference is the wider separatrix in Figure \ref{6brOM2J1630}.
In the separatrix the Lyapunov exponent is negative; thus, chaos is
still confined to the prograde region.

\begin{figure}
\plotone{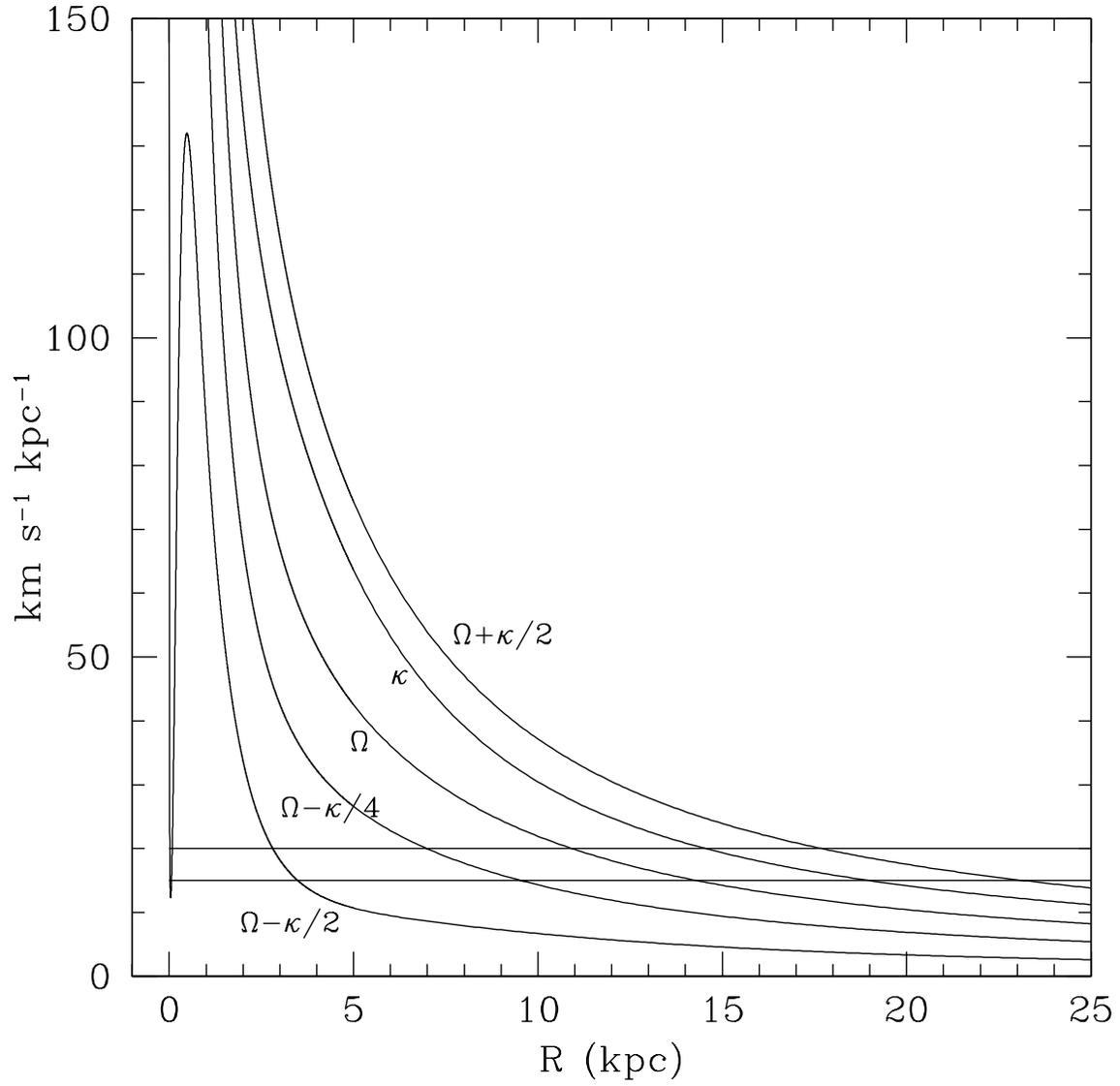}                      
\caption {Some resonance curves in the A\&S Galactic model. The lines
${\Omega}_{p}$ = 15 and 20 km s$^{-1}$~kpc$^{-1}$ are shown in the
figure.}
\label{resonancias1}
\end{figure}

\begin{figure}
\plotone{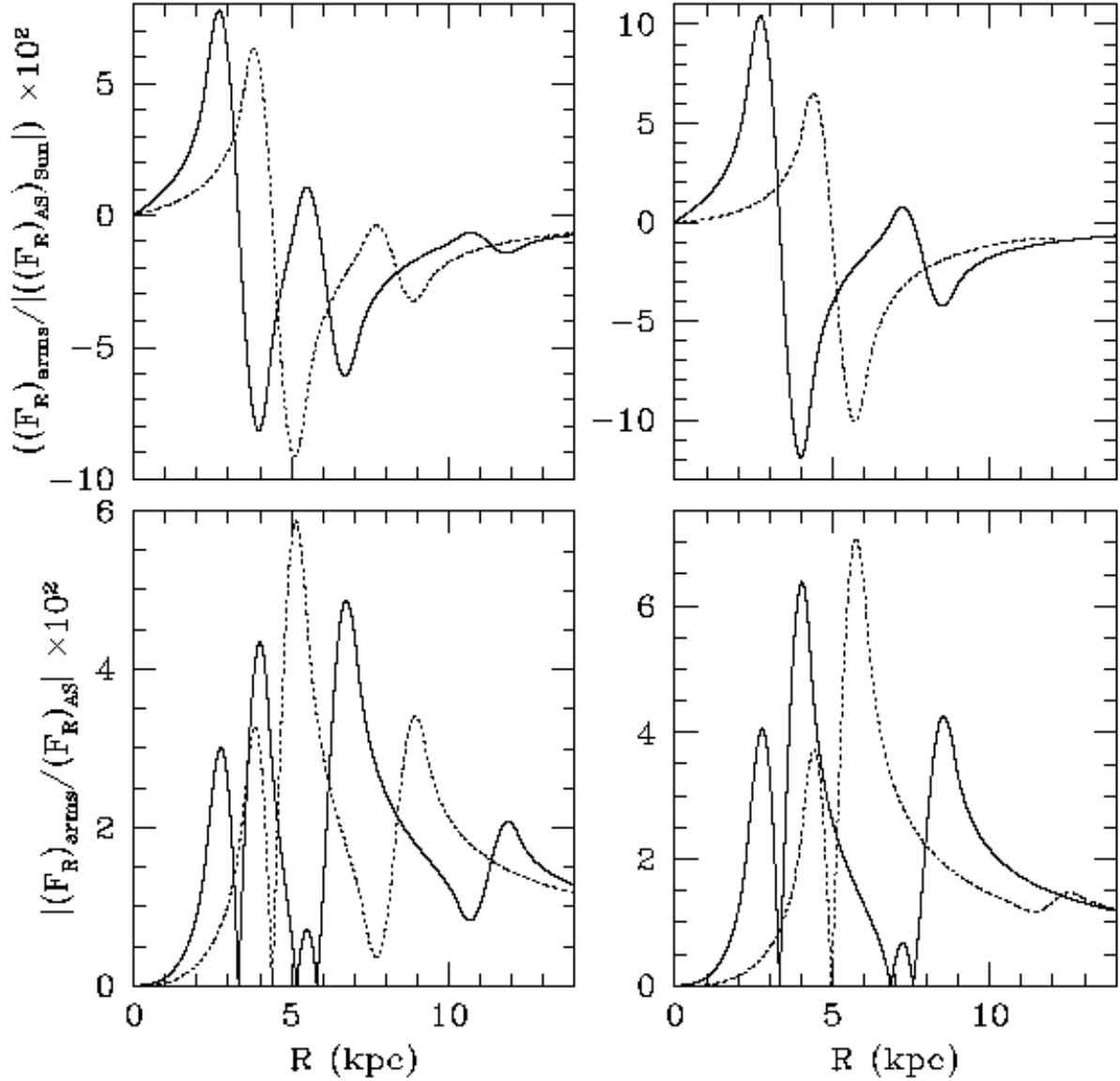}                      
\caption {Radial force due to the spiral arms scaled
by the absolute value of the A\&S force at the Solar position (upper
frames), and relative force perturbation (lower frames) in a model
with $M_S/M_D$ = 0.0175. Left-hand frames: model with the spiral locus
of Fig. \ref{locus1}; right-hand frames: model with the spiral locus
pf Fig. \ref{locus2}. Values are given along the positive $x'$
(continuous lines) and $y'$ (dotted lines) axes defined in
Fig. \ref{equipotenciales}.}
\label{fuerzas}
\end{figure}

\begin{figure}
\plotone{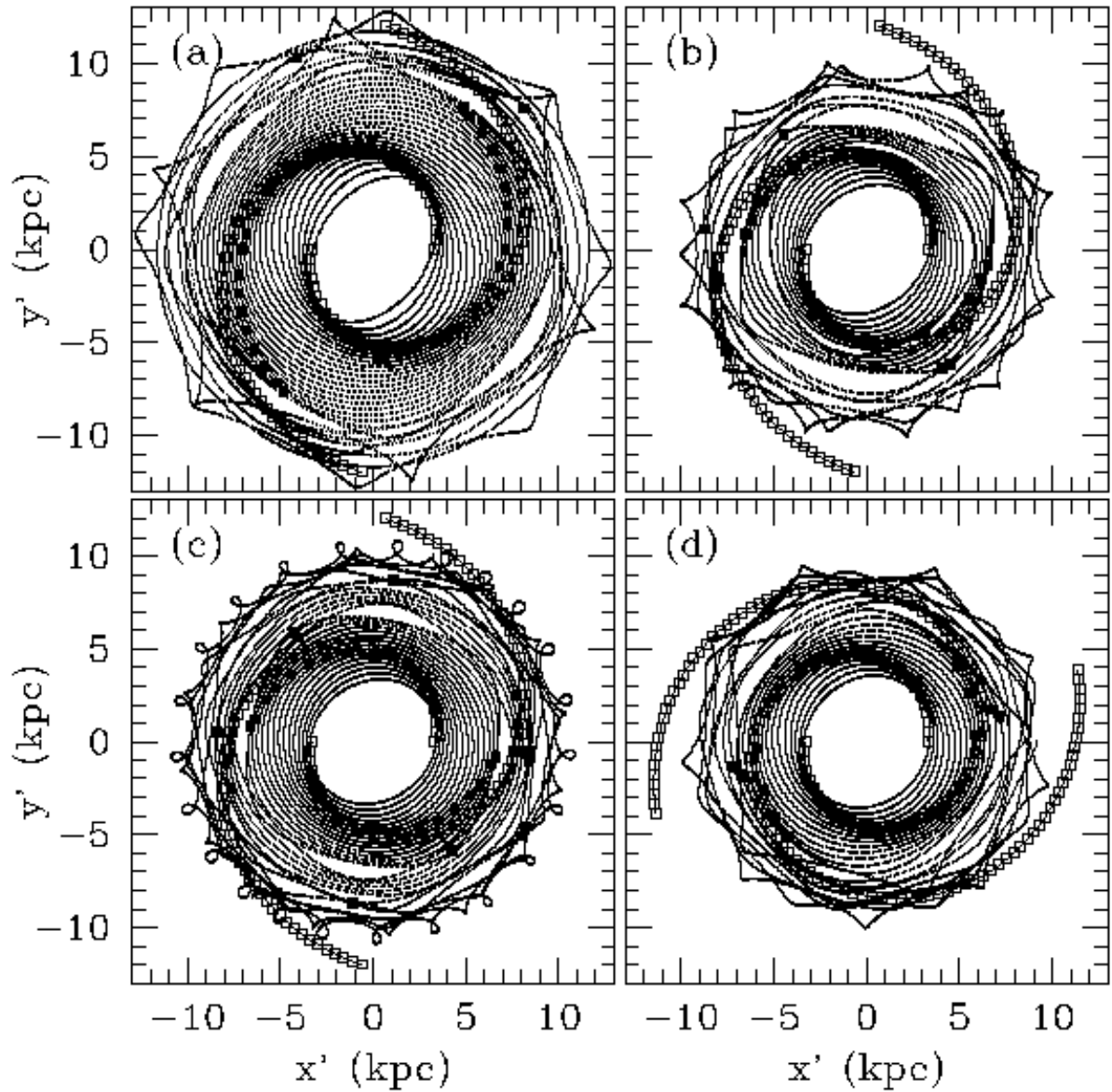}                      
\caption {Response maxima (black squares) in models with the spiral
loci (open squares) in Fig. \ref{locus2} (frames (a),(b),(c)), and in
Fig. \ref{locus1} (frame (d)). ${\Omega}_{p}$ = 15 km
s$^{-1}$~kpc$^{-1}$ in case (a); ${\Omega}_{p}$ = 20 km
s$^{-1}$~kpc$^{-1}$ in cases (b),(c),(d). $M_S/M_D$ = 0.0175 in cases
(a),(b),(d); $M_S/M_D$ = 0.00875 in case (c). Periodic orbits
calculated following the C\&G86 method are also shown.}
\label{periodicas}
\end{figure}

\begin{figure}
\plotone{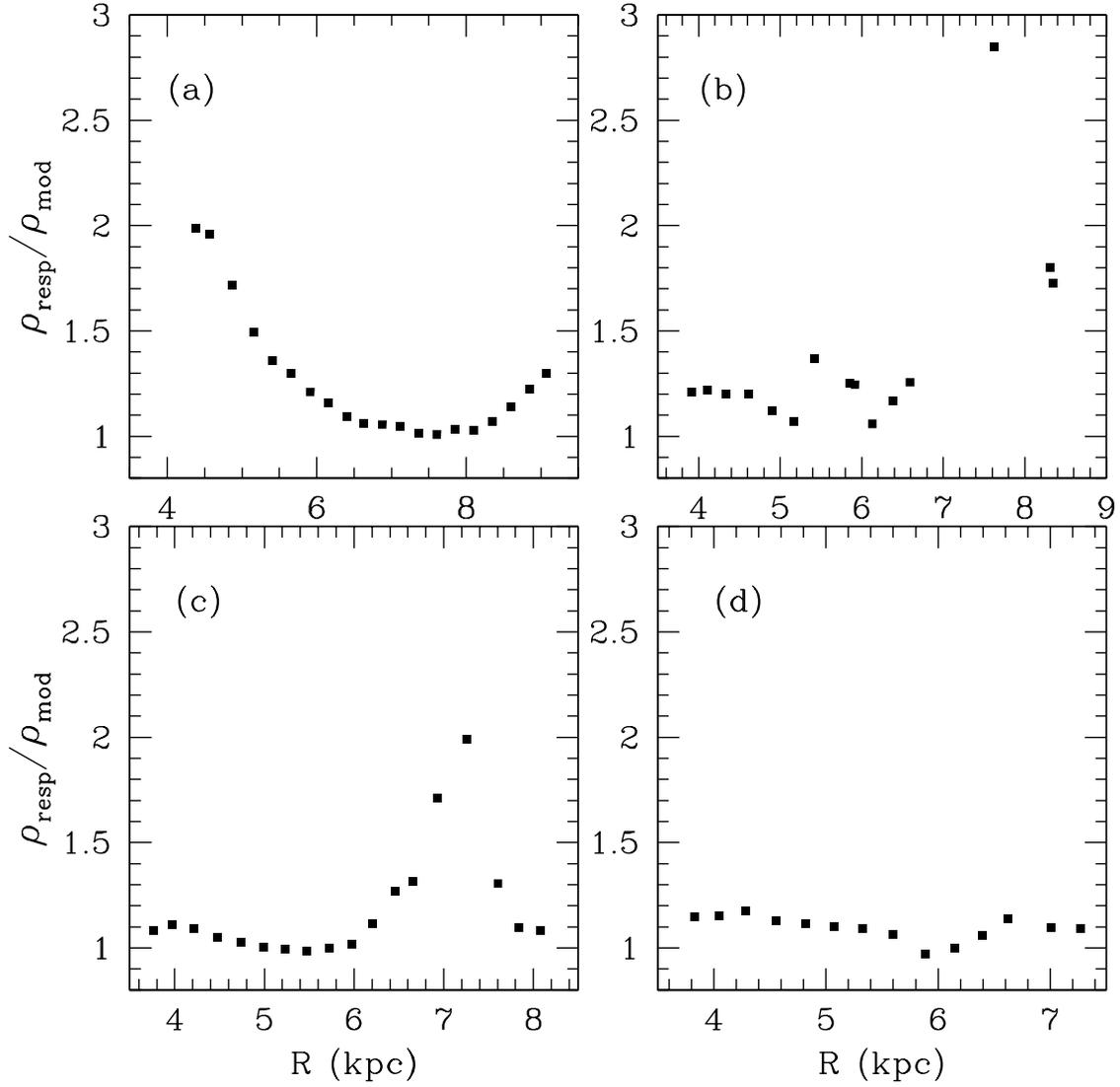}                      
\caption {The ratio of the response density to the imposed density along the arm,
${\rho}_{resp}/{\rho}_{mod}$, as a function of galactocentric
distance, for each case in Fig. \ref{periodicas}.}
\label{contraste}
\end{figure}

\begin{figure}
\plotone{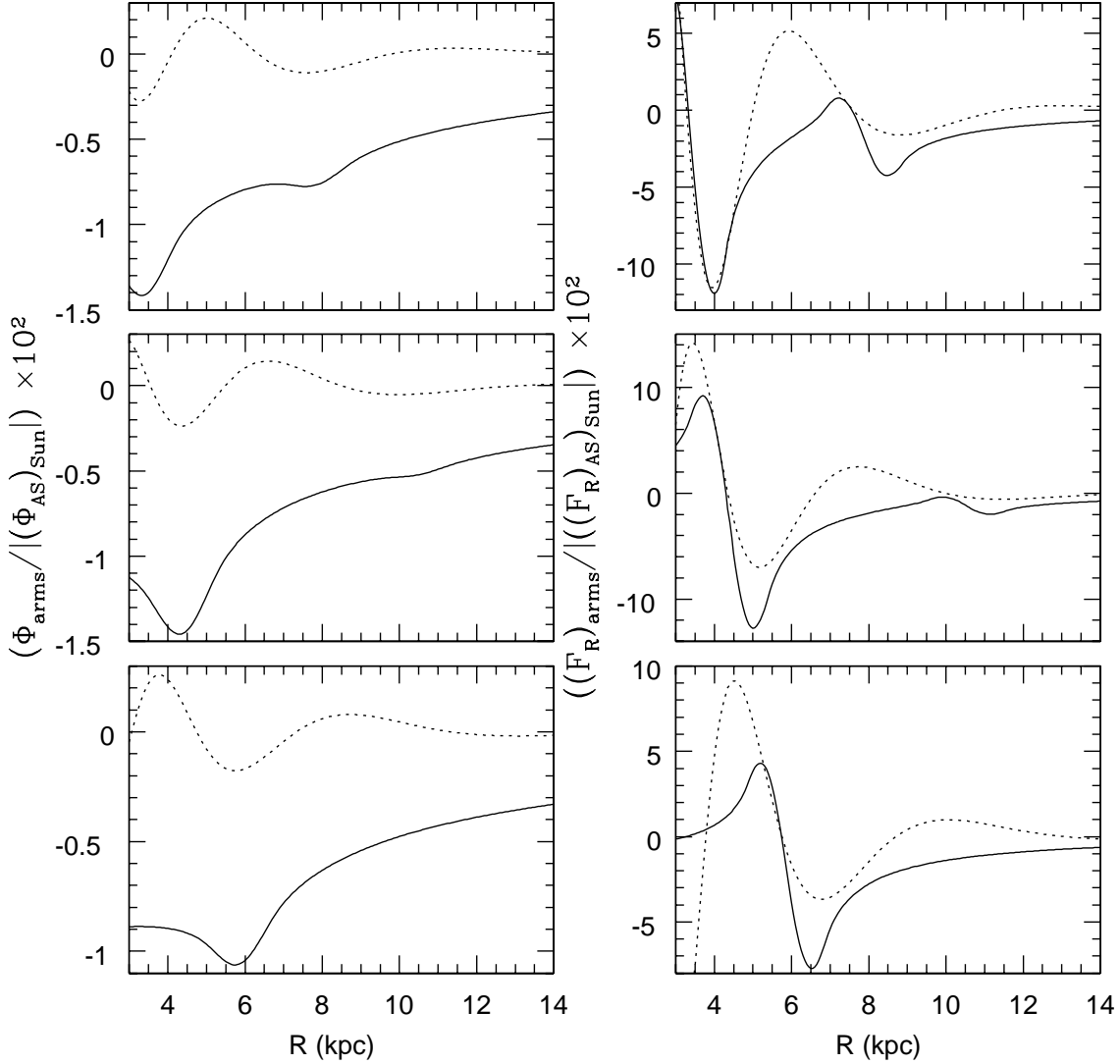}                      
\caption {Potential and radial force, scaled by the absolute value  of
the potential and force of the A\&S model at the Solar position, of a
model (continuous line) with the spiral locus in Fig. \ref{locus2}, a
mass ratio $M_S/M_D$ = 0.0175, and an exponential fall (in the central
density of the spheroids along the spiral arm) with scalelength of 2.5
kpc. The potential and radial force of the spiral arms along three
radial lines are given, one along the positive $x'$ axis (upper
frames), and the other two along the lines at 60$^\circ$ (middle
frames) and 120$^\circ$ (lower frames) from the $x'$ axis (in the
direction toward the $y'$ axis). The dotted lines show the
corresponding potential and radial force of a tight-winding model,
with the same spiral locus as in our model of Fig. \ref{locus1}.}
\label{TW1}
\end{figure}

\begin{figure}
\plotone{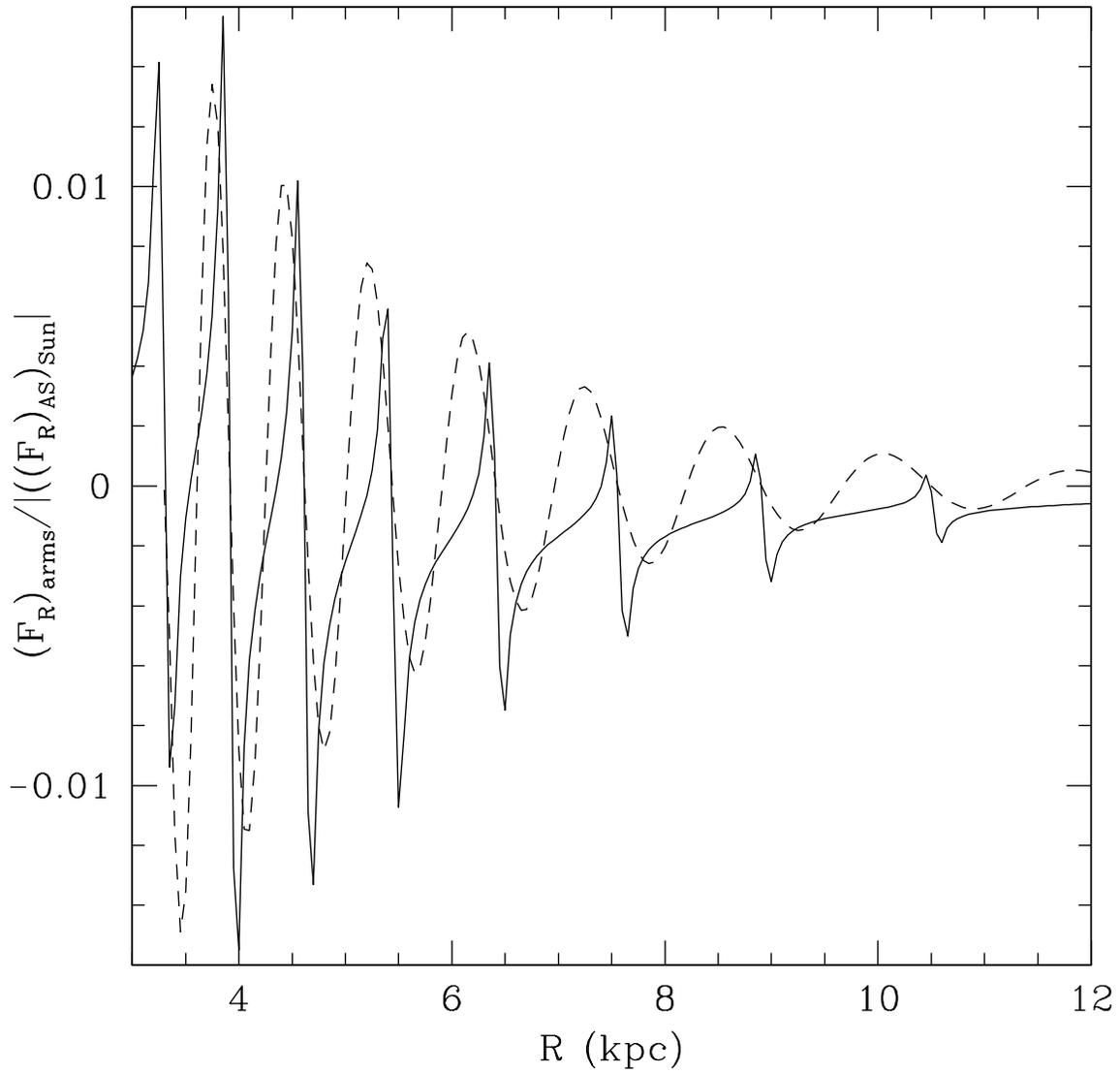}                      
\caption {Comparison of the radial forces due to a 3$^\circ$
pitch-angle, two-armed, low-mass, spiral perturbation. Our model is
given by the continuous line; the dashed line represents a
tight-winding model.}
\label{TW2}
\end{figure}

\begin{figure}
\plotone{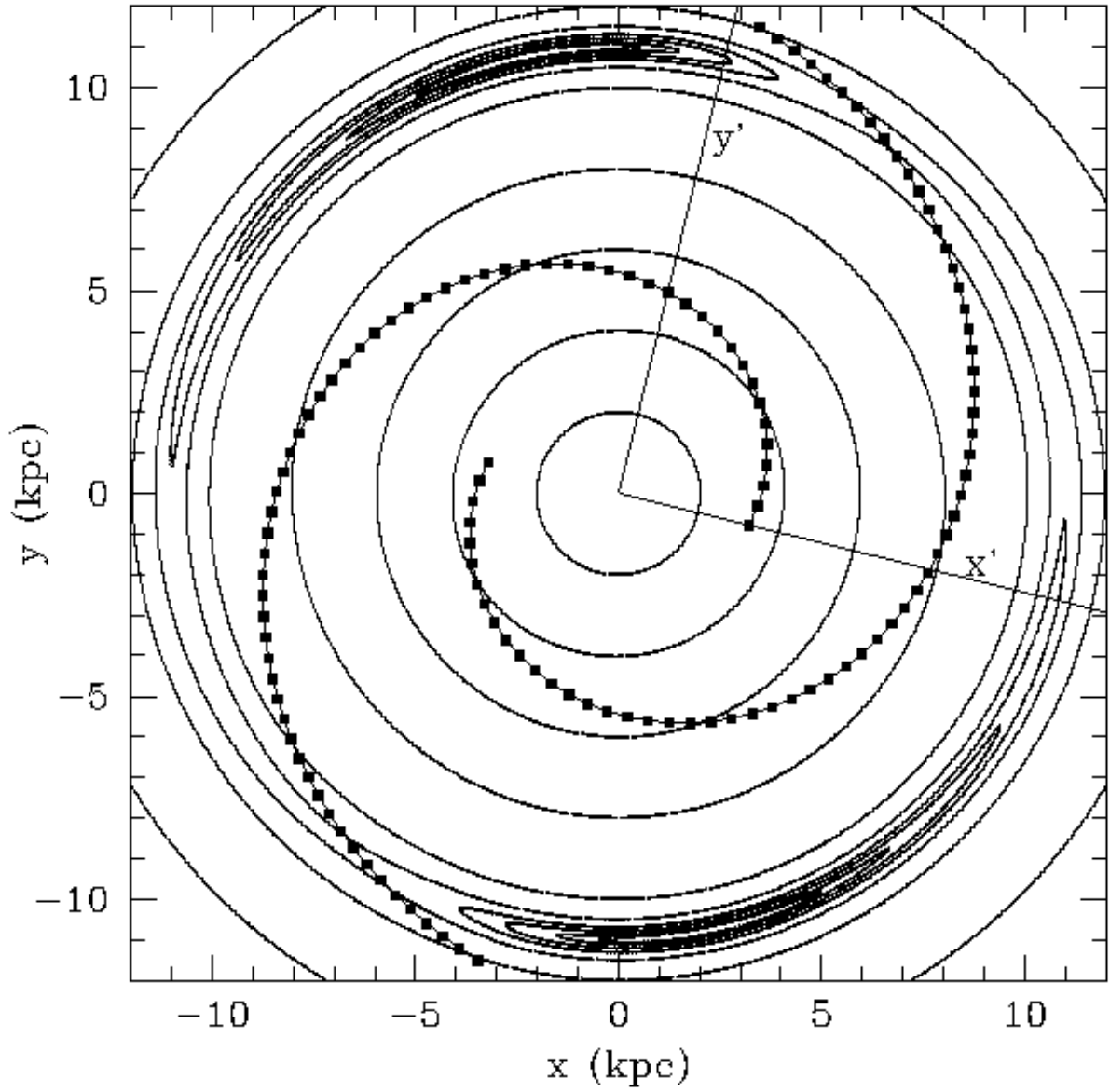}                      
\caption {Some equipotential curves ${\Phi}_{eff}$ = const. for a
model with $M_S/M_D = 0.0175$, ${\Omega}_{p}$ = 20 km
s$^{-1}$~kpc$^{-1}$, and the spiral locus in Figure \ref{locus2}. The
inertial $x,y$ and non-inertial $x',y'$ axes are shown. Each square
marks the center of an oblate spheroid.}
\label{equipotenciales}
\end{figure}

\begin{figure}
\plotone{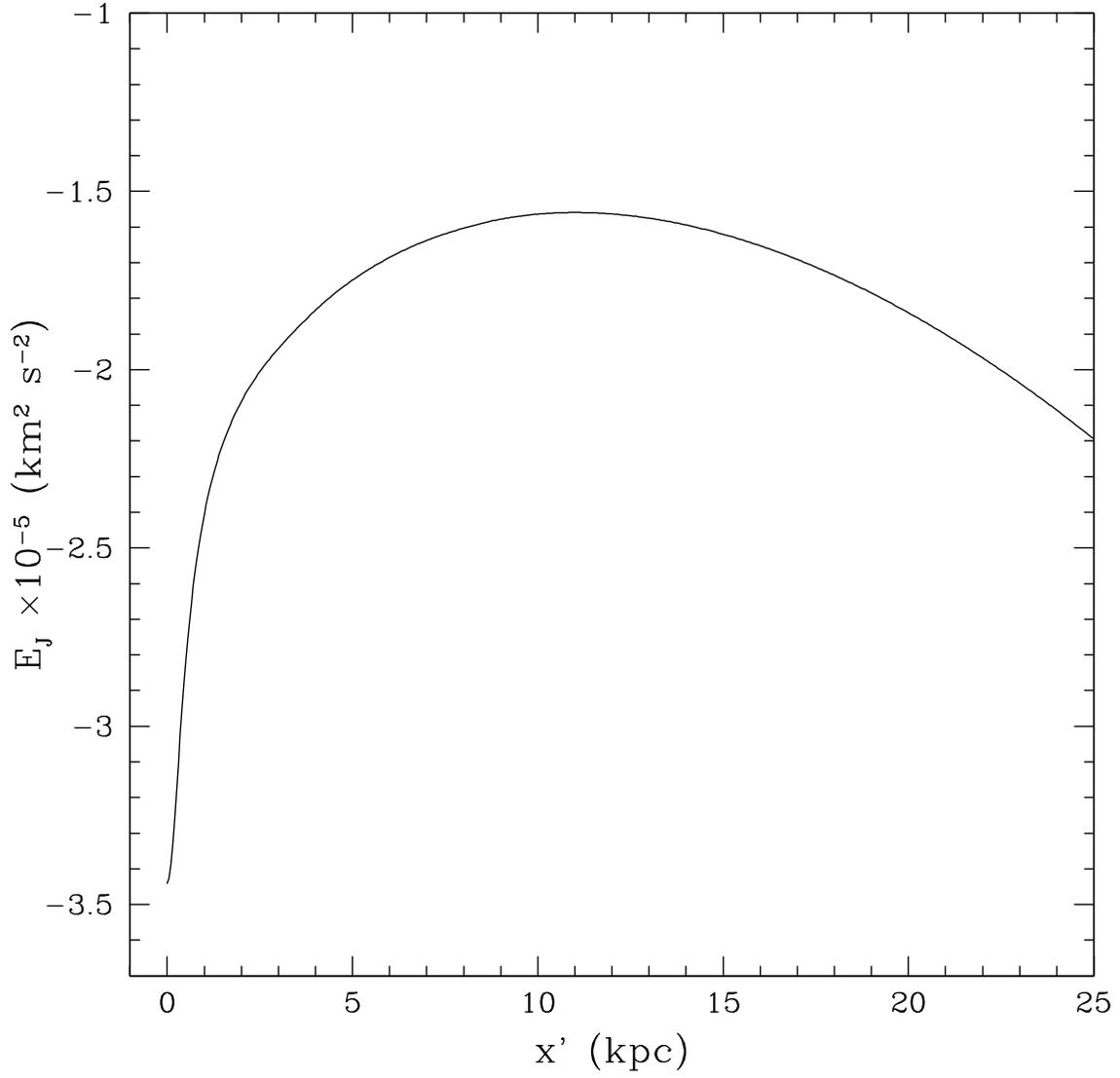}                      
\caption {The value of $E_J$ along the $x'$ axis for our model in
Fig. \ref{equipotenciales}.}
\label{Jvsx}
\end{figure}

\begin{figure}
\plotone{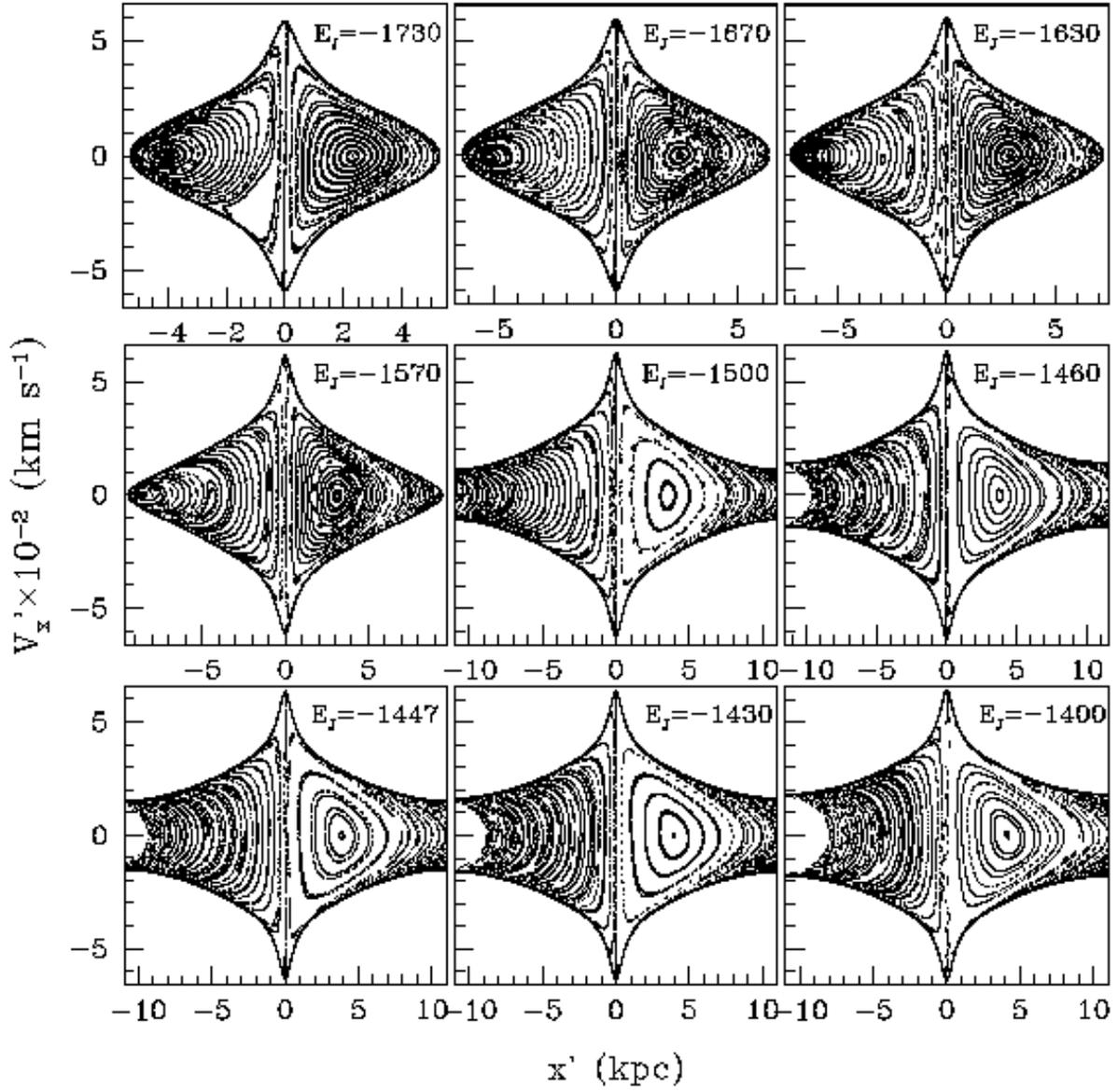}                      
\caption {Poincar\'e diagrams for nine values of $E_J$ (in units of
$10^2$~${\rm km}^2 ~{\rm s}^{-2}$), in a model with the spiral locus
of Fig. \ref{locus2}, $M_S/M_D= 0.0175$, and ${\Omega}_{p}$ = 20 km
s$^{-1}$~kpc$^{-1}$. In each plot the separatrix is shown with darker
spots.}
\label{poincare9}
\end{figure}

\begin{figure}
\plotone{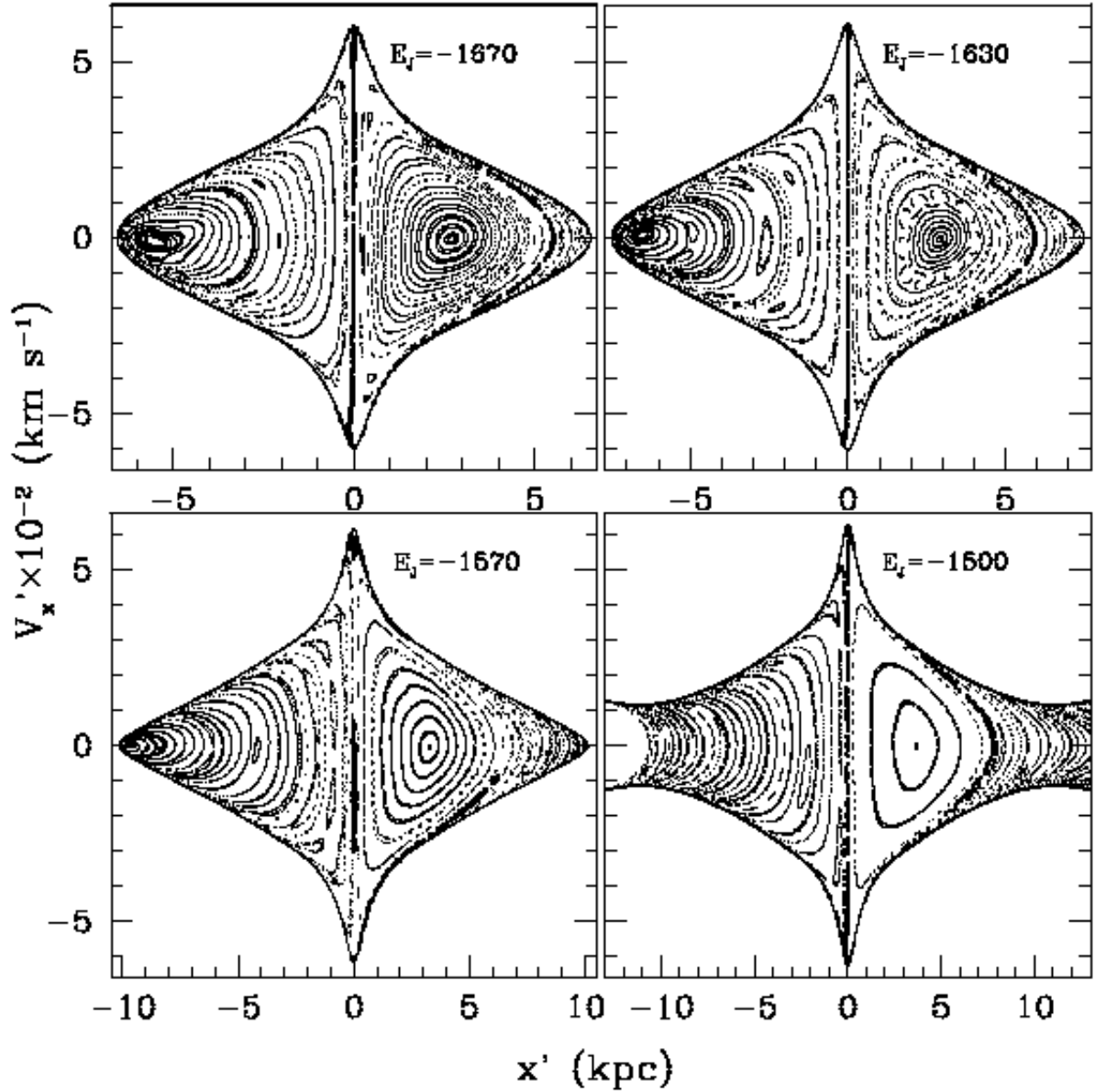}                      
\caption {Poincar\'e diagrams in a model with the six spiral arms in
Fig. \ref{locus3}, and $\Omega_p = 20$ km s$^{-1}$ kpc$^{-1}$. The two
K-band arms and the four optical arms have the same mass ratio
$M_S/M_D= 0.0175$. The separatrix is shown with darker spots.}
\label{6brazosOMP2}
\end{figure}

\begin{figure}
\plotone{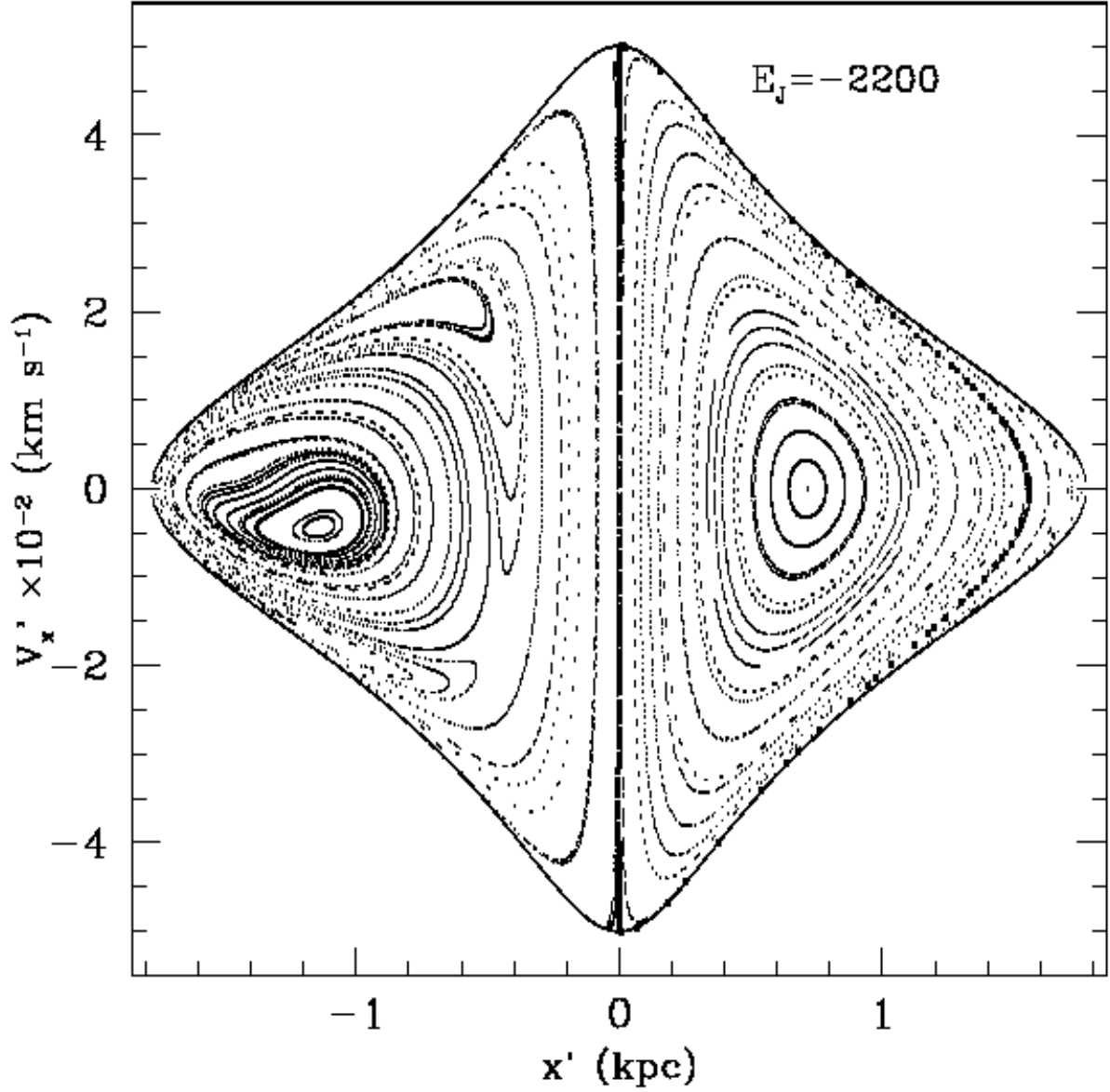}                      
\caption {Poincar\'e diagram with $E_J = -2200 \times 10^2$
km$^2$~s$^{-2}$ in the model of six spiral arms and same parameters as
in Fig. \ref{6brazosOMP2} but here with $\Omega_p = 60$ km
s$^{-1}$~kpc$^{-1}$.}
\label{MSD0.05_J2200}
\end{figure}

\begin{figure}
\plotone{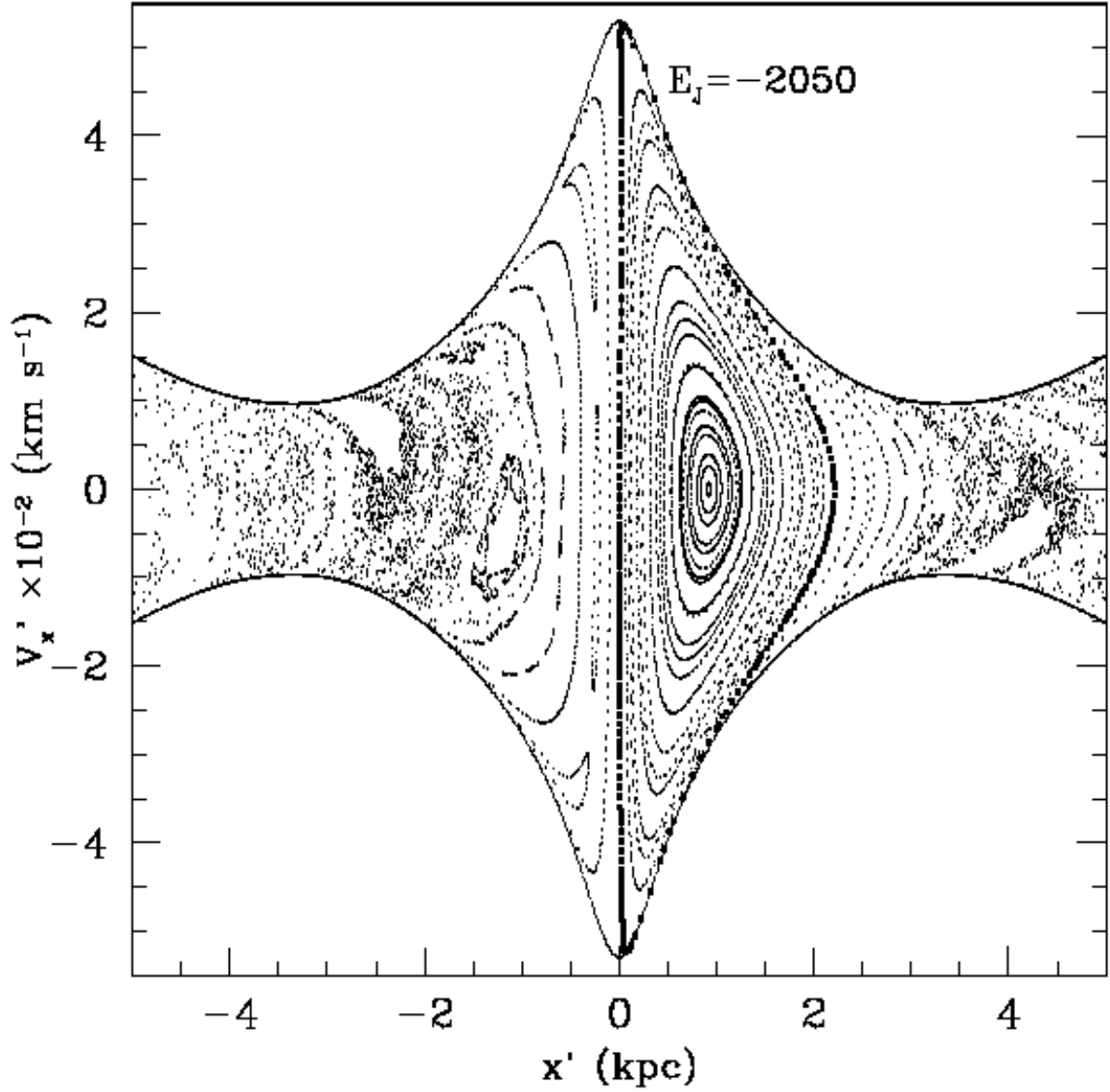}                      
\caption {Poincar\'e diagram with $E_J = -2050 \times 10^2$
km$^2$~s$^{-2}$ and same conditions as in Fig. \ref{MSD0.05_J2200}.}
\label{MSD0.05_J2050}
\end{figure}

\begin{figure}
\plotone{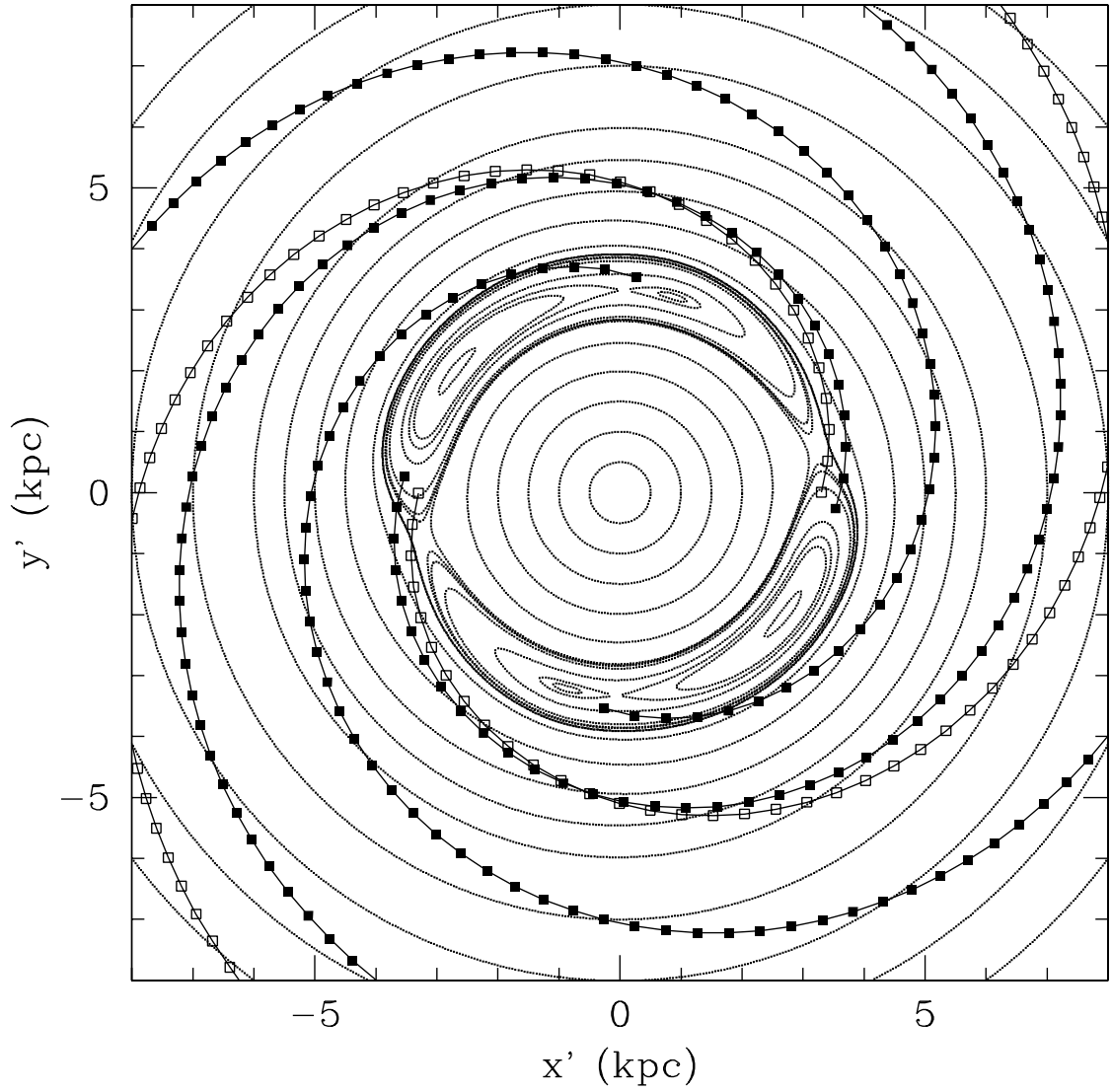}                      
\caption {Some equipotential curves ${\Phi}_{eff}$ = const.  for the
model corresponding to Figs. \ref{MSD0.05_J2200} and
\ref{MSD0.05_J2050}. Black squares: optical arms; open squares:
K-band arms.}
\label{equipotenciales6}
\end{figure}

\begin{figure}
\plotone{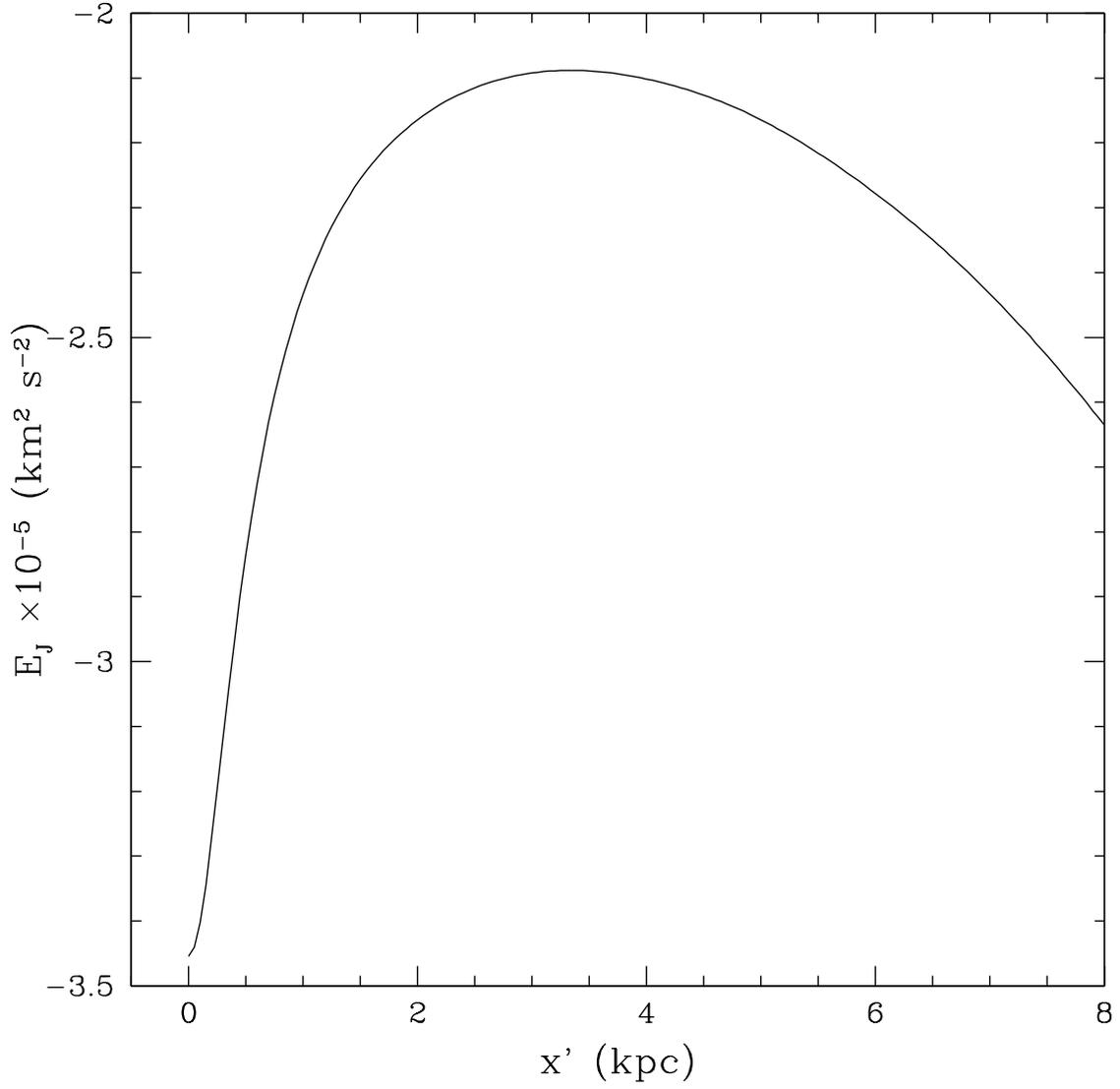}                      
\caption {The value of $E_J$ along the $x'$ axis in
Fig. \ref{equipotenciales6}.}
\label{Jvsx6}
\end{figure}

\begin{figure}
\plotone{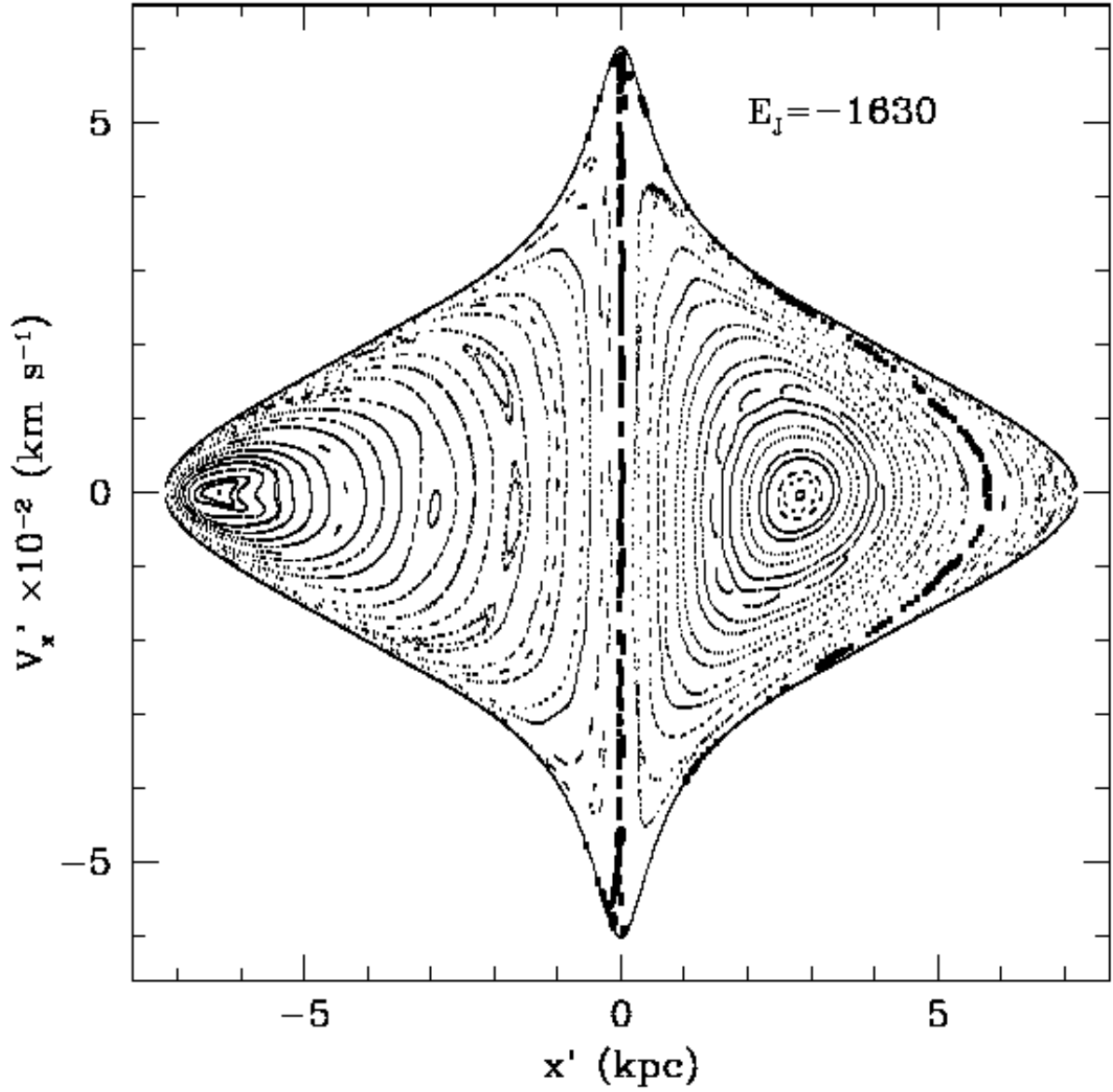}                      
\caption {Zoom of the Poincar\'e diagram with $E_J = -1630 \times 10^2$ 
km$^2$~s$^{-2}$ in Fig. \ref{poincare9}. Compare with next figure.}
\label{MSD0.0175}
\end{figure}

\begin{figure}
\plotone{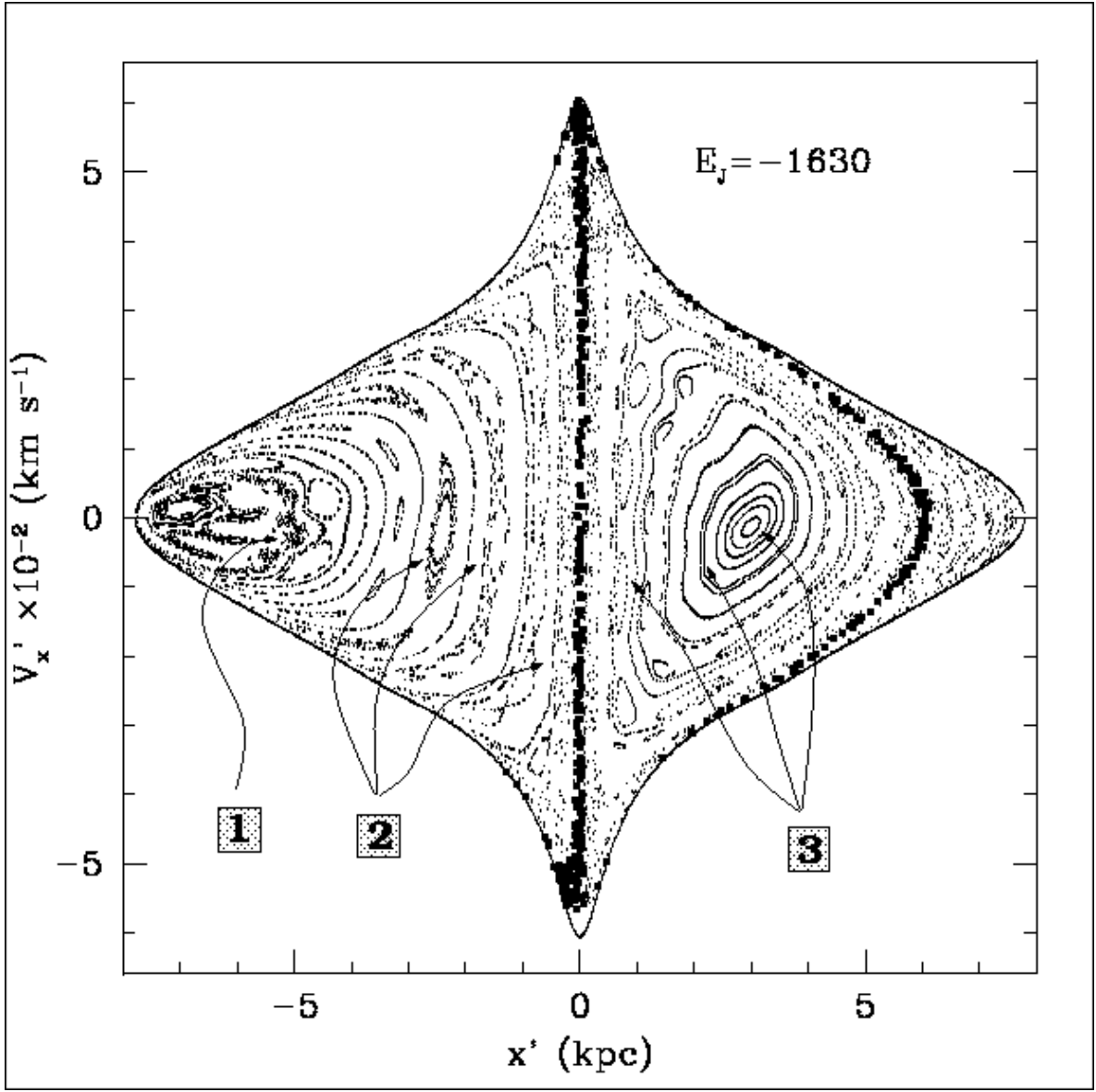}                      
\caption {Poincar\'e diagram for the same value of $E_J$ and same
parameters as in Fig. \ref{MSD0.0175}, but here with $M_S/M_D =
0.05$. The arrows and numbers indicate some zones where we computed
Lyapunov exponents.}
\label{MSD0.05}
\end{figure}

\begin{figure}
\plotone{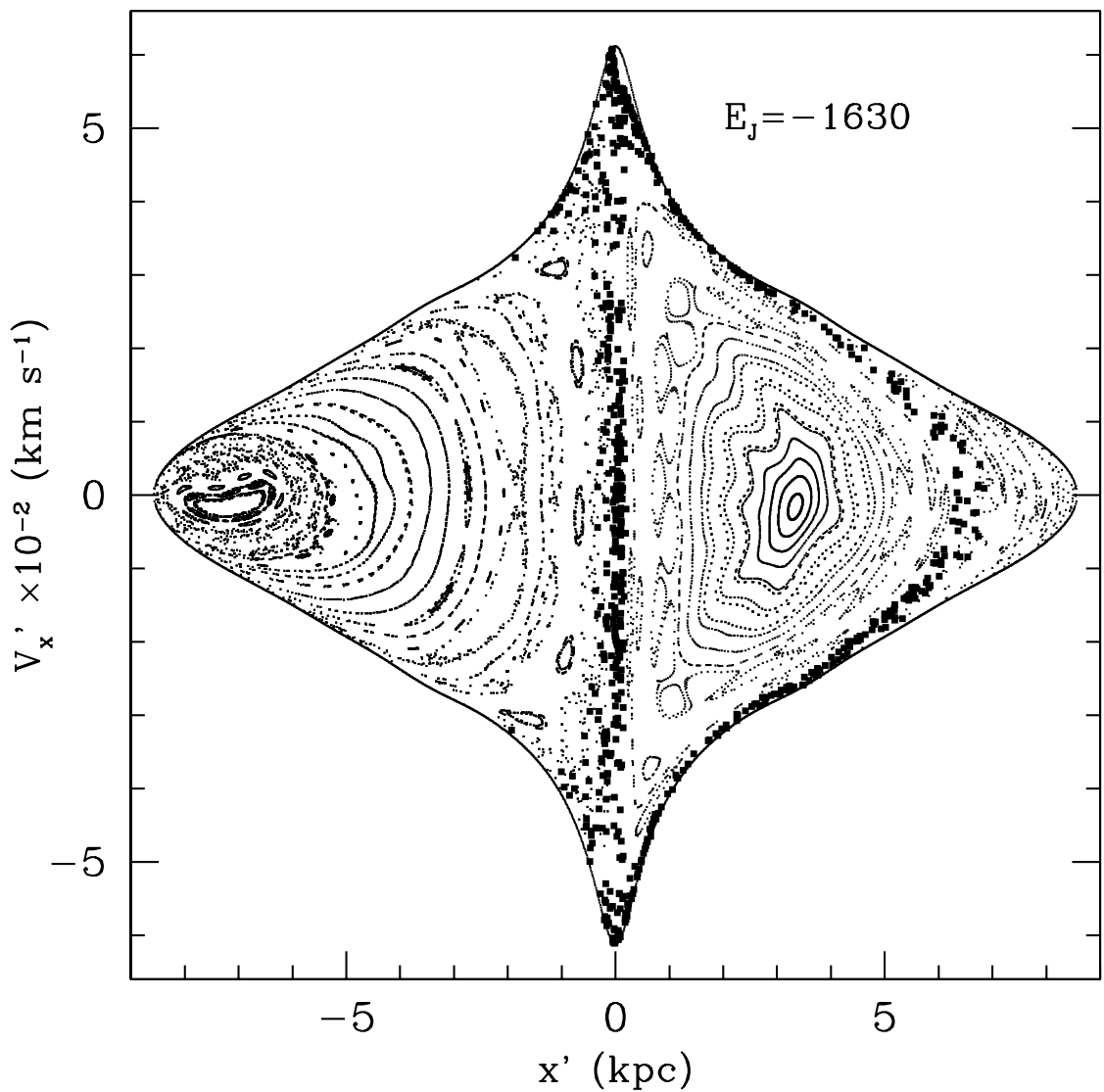}                      
\caption {Poincar\'e diagram with $E_J = -1630 \times 10^2$
km$^2$~s$^{-2}$ in a model with the six spiral arms in
Fig. \ref{locus3}. $\Omega_p = 20$ km s$^{-1}$ kpc$^{-1}$, and
$M_S/M_D = 0.05$ for both optical and K-band arms. Compare with
Fig. \ref{MSD0.05}, which has the same ${\Omega}_{p}$ and $M_S/M_D$
in a model with the spiral locus in Fig. \ref{locus2}.}
\label{6brOM2J1630}
\end{figure}

\begin{figure}
\plotone{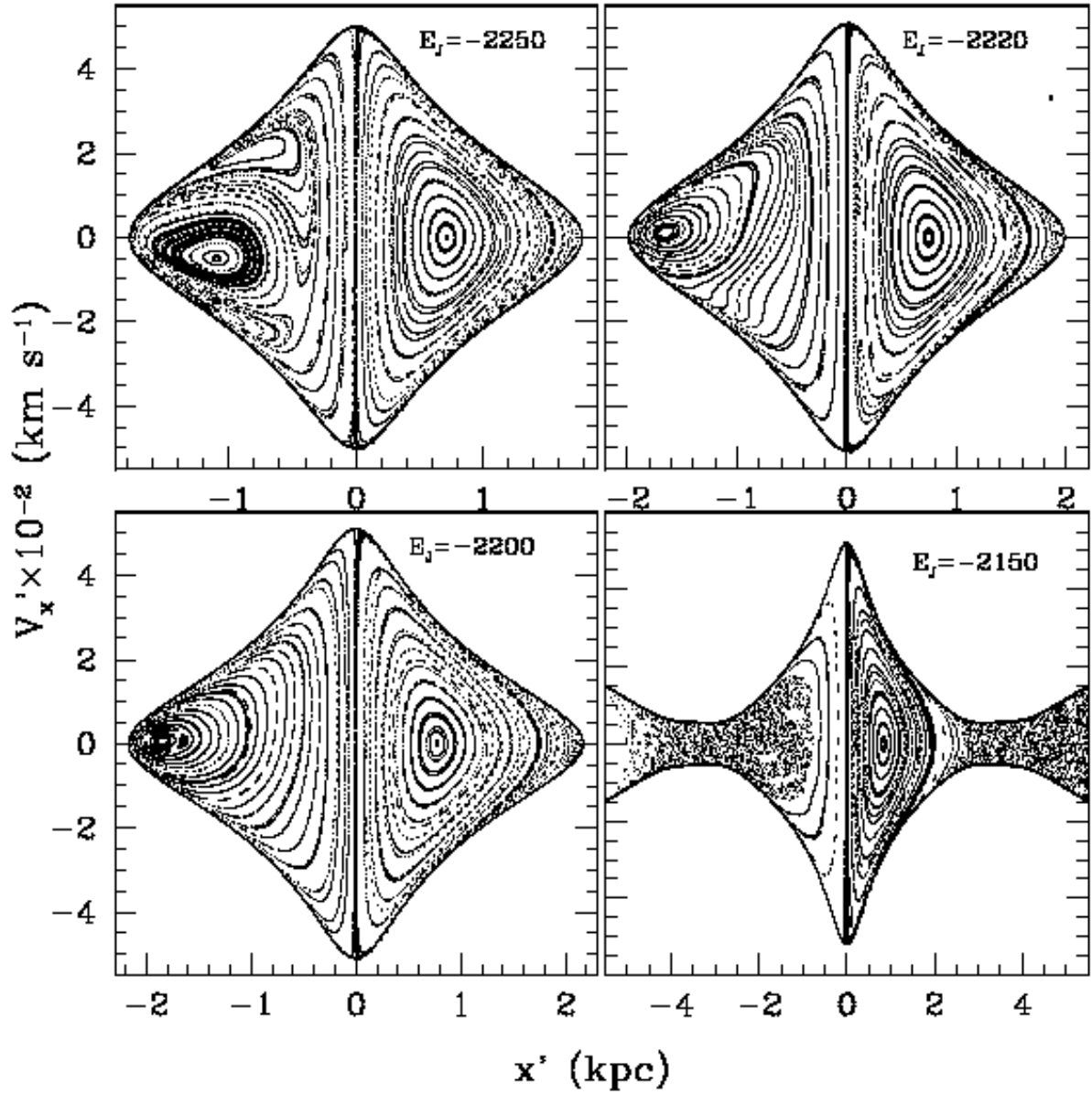}                      
\caption {Some Poincar\'e diagrams in a model with the six spiral arms
in Fig. \ref{locus3}. $\Omega_p = 60$ km s$^{-1}$ kpc$^{-1}$, and
$M_S/M_D = 0.05$ for both optical and K-band arms. The separatrix is
shown with darker spots.}
\label{6brazosOMP6}
\end{figure}

In Figure \ref{6brazosOMP6} we give four Poincar\'e diagrams for the
model with six spiral arms in Figure \ref{6brOM2J1630}, but now with
${\Omega}_{p}$ = 60 km s$^{-1}$ kpc$^{-1}$. The prograde and
retrograde regions are separated by a narrow separatrix. The first
three diagrams show non-chaotic motion within corotation distance (see
Figures \ref{resonancias1} and \ref{equipotenciales6}). The fourth
diagram, in the lower right frame, shows pervasive chaos in a wide
zone in the prograde region. This corresponds to stellar motion that
can surpass the corotation barrier. The same behavior is obtained in
the case shown in Figure \ref{MSD0.05_J2050}, with a lower mass in the
six spiral arms. This transition from regular to chaotic motion was
also obtained by Fux (2001), who considered the Galactic bar as the
perturbating agent.

\begin{figure}
\plotone{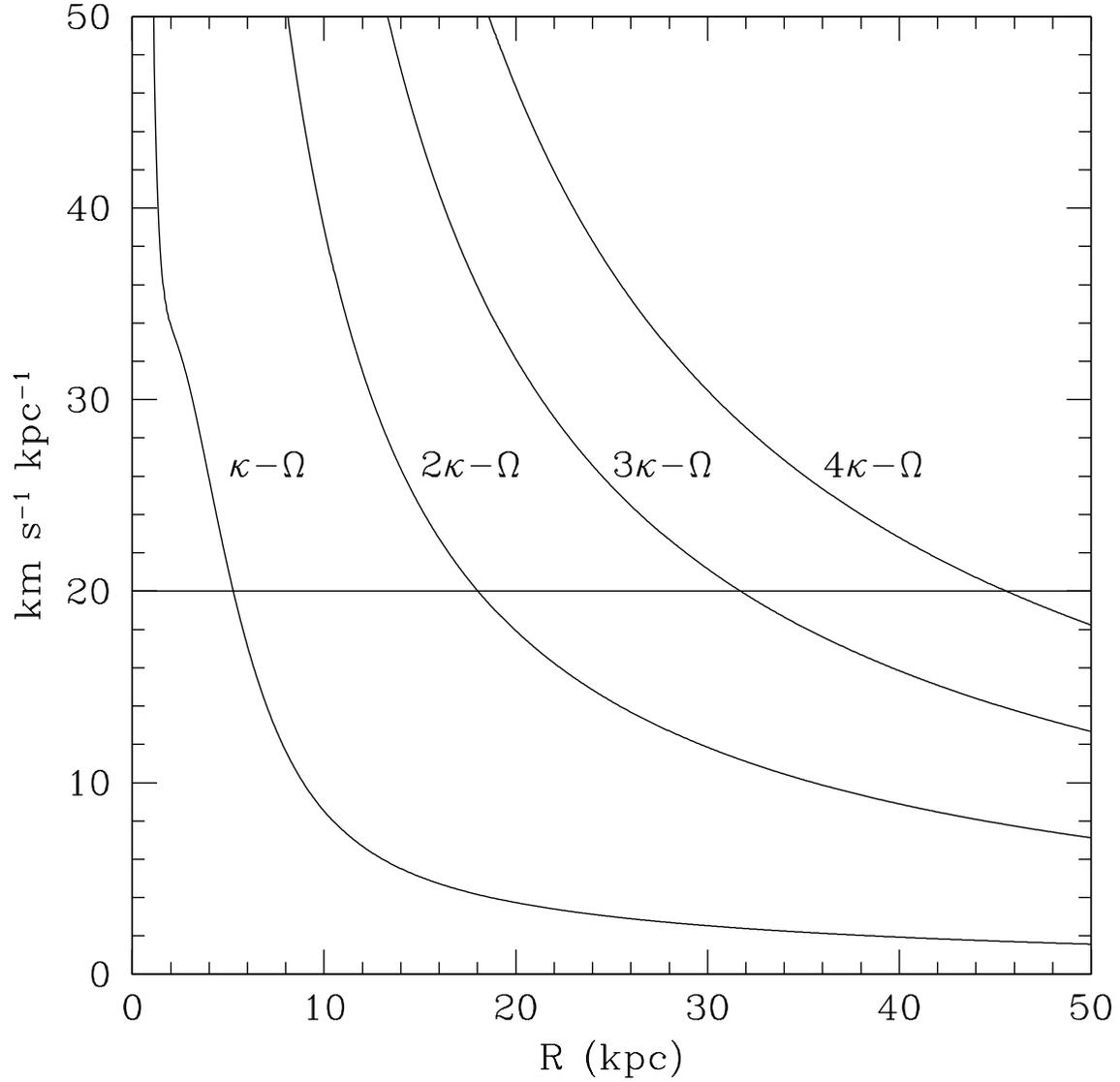}                      
\caption {Some resonance curves for nearly circular retrograde orbits
in the A\&S Galactic model. The line $\Omega_p = 20$ km s$^{-1}$
kpc$^{-1}$ is given in the figure. Compare with
Fig. \ref{resonancias1}, which corresponds to nearly circular
$prograde$ orbits.}
\label{resonancias2}
\end{figure}

All these results show that the onset of chaotic motion starts in the
prograde region. The prime cause for chaotic motion is the onset of
bifurcations and resonance interactions (Contopoulos 1967, Martinet
1974, Athanassoula et al. 1983, Athanassoula 2001). Regarding the
resonance interactions, Figure \ref{resonancias2} shows some important
resonance curves for nearly circular retrograde orbits in the A\&S 
Galactic model; in
particular, with the ${\Omega}_{p}$ = 20 km s$^{-1}$ kpc$^{-1}$ line
given in the figure (corresponding to the pattern speed of a spiral 
perturbation rotating in the {\it prograde} sense), the corresponding 
resonance positions can be read
on the R axis. Thus, resonances for retrograde orbits are more widely
separated, as compared with the resonances for prograde orbits, some
of which are shown in Figure \ref{resonancias1}; i.e., resonance
interactions are more important in prograde orbits.

\section{DISCUSSION AND CONCLUSIONS}

We present a 3D model for a spiral mass distribution, consisting of
inhomogeneous oblate spheroids superposed along a given spiral
locus. The model is applied in particular to our Galaxy, but can
easily be applied to spiral galaxies in general. Furthermore, it
allows to look with a deeper physical insight into details that are
inaccessible to the classical treatment of the spiral perturbation,
which models it as a simple periodic function.  Our model of oblate
spheroids is physically simple and plausible, with continuous
derivatives and density laws.

In our Galaxy, the model parameters, such as the number of spiral
arms, its pitch angle, its radial extent, the pattern speed, the
dimensions and mass density of the spheroids, and the total mass in
the arms, were taken in a range of possibilities suggested by
observations and theory.

In principle, the dimensions and mass density of the oblate spheroids
will depend on the type of spiral arms which are modeled, gaseous or
stellar. In this first work the adopted dimensions resemble those of
gaseous spiral arms (Kennicutt \& Hodge 1982), and a linear density
law in the spheroids has been considered. We assembled Galactic models
with two-armed spirals, such as the 15.5$^\circ$ pitch-angle, stellar
arms discussed by Drimmel (2000), and with $six$ spiral arms: adding
the four 12$^\circ$ pitch-angle, optical arms, delineated by luminous
HII regions. From a range of possibilities, we considered three values
of the pattern speed: ${\Omega}_{p}$ = 15, 20, and 60 km s$^{-1}$
kpc$^{-1}$, and the ratio of the mass in the spiral arms to the disk's
mass in the A\&S axisymmetric Galactic model, $M_S/M_D$, in the range
0.0175 to 0.05 . In this range of masses the average force due to the
spiral arms is between 5 and 10 $\%$ of the background axisymmetric
force.

In an effort to achieve a self-consistent model of the spiral
perturbation in our Galaxy, we have used the well-known, approximate
method of C\&G86 to analyze the density response to this imposed
perturbation. We have computed the density response in a Galactic
potential with two spiral arms, taking the pattern speed as
${\Omega}_{p}$ = 15 and 20 km s$^{-1}$ kpc$^{-1}$, and the mass ratio
$M_S/M_D$ around the lower limit given above. Our nearly
self-consistent models favor the pattern speed of 20 km s$^{-1}$
kpc$^{-1}$. However, this preliminary analysis must be improved at
least accounting for (a) a hot stellar population around the central
periodic orbits, (b) a four-armed, stellar, spiral pattern in the
density response, in addition to the main two-armed component (A\&L),
and (c) a properly modeling of the dimensions of two-armed stellar
spirals, as the K-band arms given by Drimmel (2000); this type of arms
is azimuthally broad (Rix \& Zaritsky 1995), thus an increase with
galactocentric distance of the major semi-axis $a_0$ of spheroids
would be appropriate. This analysis will be presented in a future work.

Modeling of the gravitational potential produced by a spiral
perturbation has usually been based on the {\it tight-winding
approximation} (TWA, Eq. 1). We have compared the potential and force
fields of a two-armed spiral perturbation given by our model with a
TWA model. We found that the self-gravity of the spiral pattern
(i.e. contributions to the potential from the entire pattern), which
is not accounted for in the TWA (which acts more like a local
approximation), cause the local spiral potential to adopt shapes that
are not correctly fit by the simple perturbing term that has been
traditionally invoked to represent the local spiral potential. This
fact may have far-reaching consequences; for instance, in the gas
response to the spiral perturbation. We have performed modest 1D MHD
simulations (Franco, Martos \& Pichardo 2001) with the code Zeus to
show the differences in the gas response using the conventional model
of a cosine for the potential and the model presented in this
work. These simulations show that shocks do not leave the arm
downstream as in previous calculations (Baker \& Barker 1974, Martos
\& Cox 1998) for a plausible range of entry speeds. And, in
correspondence with observational expectations, shocks seek the
upstream edge of the arm, i.e. the concave side inside corotation
marked in optical observations of galaxies for accumulations of dust
in the inner part of the spiral arms. The inclusion of the magnetic
field is essential to this effect.  In this manner, results based on
the TWA should be revised: the gas response depends strongly on the
position in the Galaxy. A potential ``well'' in the arm may disappear
as such at a different segment of the arm.

In the analysis of Poincar\'e diagrams we found it is quite fruitful
to use an inertial frame to define the prograde or retrograde sense of
orbital motion around the Galactic center along with the usual
definition in the non-inertial system, where the Poincar\'e diagrams
are defined. In the inertial frame the sense of motion is preserved
with time for almost every orbit in our experiments exception being
orbits with nearly zero angular momentum. This property relies upon
the parameters we consider plausible for our Galaxy. If we include
information of the inertial system in the non-inertial one, Poincar\'e
diagrams reveal that prograde and retrograde orbits, as defined in the
inertial frame, occupy sharply separated regions, through a separatrix
corresponding, loosely, to nearly zero angular momentum orbits in this
system.

The definition of sense of orbital motion in the inertial frame goes
beyond a mere matter of semantics, for it has a simple physical
meaning and it appears to be intimately connected to the onset of
chaos. Based on an analysis of Poincar\'e diagrams and the first
Lyapunov exponent we find that, within plausible amplitudes and pitch
angles of the spiral arms for a Galaxy such as the Milky Way (and
independently of the number of arms chosen), if there is chaos, only
prograde orbits can exhibit it, and for a sufficiently weak
perturbation, as it seems to be the case in our Galaxy, the separatrix
is a well-defined narrow curve. The onset and extension of chaotic
subregions of the prograde region, depends on two main parameters that
are the mass in the spiral arms or the relative force; and the angular
velocity. We stress out the point that the standard definition in the
rotating frame, which calls the $x' > 0$ of the diagram the retrograde
side, and $x' < 0$ the prograde side (for a spiral pattern moving
clockwise), would not have shed light onto the connection
chaos-prograde motion, since the same orbit (ordered or chaotic) can
occupy both sections of Poincar\'e diagrams. The different behavior
regarding the onset of chaos of prograde and retrograde orbits, as
defined in the inertial system, could be attributed to the overlapping
of resonances (Contopoulos 1967, Athanassoula 2001). Figures
\ref{resonancias1} and
\ref{resonancias2} show that the spacing of the main resonances is
wider for retrograde orbits, and is even smaller if we take higher
angular velocities for the spiral pattern. In cases with the lower
spiral masses ($M_S/M_D \la .03$), we do not find chaos for angular
speeds of the spiral pattern lower than 20 km s$^{-1}$ kpc$^{-1}$. We
have also computed some orbital families with $\Omega_p = 60$ km
s$^{-1}$ kpc$^{-1}$ since n-body models predict thoses velocities
(that corresponds to the bar), and we find that, even for the lowest
spiral mass we considered, chaos appears for some families
(Fig. \ref{6brazosOMP6}, $E_J = -2150 \times 10^2$ km$^2$ s$^{-2}$),
where almost all the prograde region is chaotic but chaos do not
invade retrograde region. The inclusion of more than two spiral arms
does not seem change dramatically the results (Figs. \ref{MSD0.05} for
two arms and \ref{6brOM2J1630} for 6 arms). A minimum strength of the
perturbation is required for the appearance of stochastic motion in
the models with the lowest angular speeds (15 and 20 km s$^{-1}$
kpc$^{-1}$). We find that the amplitude of approximately 6$\%$ (in
average) of the axisymmetric radial force is required (that
corresponds in our model to a $M_S/M_D \la .05$ for a pitch angle of
15.5$^\circ$). For cases of very strong spiral perturbations (relative
forces higher than 15$\%$) the separatrix is no longer a well-defined
curve and chaos is pervasive. However we don't think this spiral
forcing is proper for a Sb galaxy.

It is worth noticing that our results are valid in the plausible range
of parameters (and even in unrealistic cases with maximum relative
forces for the spiral arms up to 15$\%$) in a galaxies similar to the
Milky Way, with 2, 4 or 6 arms. However, our results will surely be
altered by the influence of the Galactic bar. We are currently
studying this effect.

 \section*{Acknowledgments}

This work was partially supported by Universidad Nacional Aut\'onoma
de M\'exico (UNAM) under DGAPA-PAPIIT grants IN130698 and IN114001.
Calculations were carried out using the Origin Silicon Graphics
supercomputer of DGSCA-UNAM.  We thank INAOE in Tonatzintla, M\'exico,
for hosting the Guillermo Haro International Workshop 2001 (and very
specially to our host, I. Puerari), where many discussions enriched
this paper, and conversations with: P. Grosb{\o}l, E. Athanassoula,
M. Bureau, W. Dehnen, P. Teuben, J. Franco, H. Dottori, E. Alfaro and
F. Masset. We are also indebted to comments from L. Sparke, D. Cox,
G. G\'omez-Reyes, C. Allen, I. King and S. Kurtz.

\clearpage

\clearpage

\clearpage

\clearpage

\clearpage

\clearpage

\clearpage

\clearpage

\clearpage

\clearpage

\clearpage

\clearpage

\clearpage

\clearpage

\clearpage

\clearpage

\clearpage

\clearpage

\clearpage

\clearpage

\clearpage

\clearpage

%\newpage

%\label{lastpage}

\end{document}